\documentclass{IEEEtaes}
\usepackage{color,array,amsthm}
\usepackage{graphicx}
\usepackage{mathrsfs}
\usepackage{diagbox}
\usepackage{optidef}
\usepackage{amsthm, amssymb}

\theoremstyle{definition}

\usepackage{amssymb}% http://ctan.org/pkg/amssymb
\usepackage{pifont}% http://ctan.org/pkg/pifont

\usepackage{multirow}
\usepackage{gensymb}
\usepackage{blindtext}
\usepackage{tabularx}
\usepackage{graphicx} % if you want to include jpeg or pdf pictures
\usepackage{bm}
\usepackage{mathtools, cuted}
\usepackage{scalerel}
\usepackage{mwe}
\usepackage{cite, balance}
\usepackage[utf8]{inputenc}

\usepackage{lipsum, color}

\usepackage{amssymb}
\usepackage{amsthm}
\usepackage{upgreek}
\usepackage{algorithm}
\usepackage{algorithmic}
\usepackage{bbold}
\usepackage{dsfont}
\usepackage[english]{babel}
\usepackage[autostyle]{csquotes}
\usepackage{bigints}
\graphicspath{ {images/} }
\usepackage{float}
\usepackage{titlesec}
\titleclass{\subsubsubsection}{straight}[\subsection]
\usepackage[numbers,sort&compress]{natbib}
\usepackage[labelfont=bf]{caption}
\usepackage{caption}
\usepackage[justification=centering]{subcaption}
\usepackage{cool}
\usepackage{booktabs}
\usepackage[english]{babel}
\usepackage[usenames,dvipsnames,svgnames,table]{xcolor}

\usepackage{caption}
\newcommand{\db}[1]{{\textcolor{black}{#1}}}

\begin{document}

\title{On-Demand Routing in LEO Mega-Constellations with Dynamic Laser Inter-Satellite Links}

\author{DHIRAJ BHATTACHARJEE}
\member{Member,~IEEE}
\author{PABLO G. MADOERY}
\member{Senior Member,~IEEE}
\author{AIZAZ U. CHAUDHRY}
\member{Senior Member,~IEEE}
\author{HALIM YANIKOMEROGLU}
\member{Fellow,~IEEE}
\affil{Carleton University, Canada} 
\author{G{\"{U}}NE{\c{S}}~KARABULUT KURT}
\member{Senior Member,~IEEE}
\affil{Polytechnique Montreal, Canada} 
\author{PENG HU}
\member{Senior Member,~IEEE}
\affil{National Research Council Canada, Canada}
\author{KHALED AHMED}
\affil{Satellite Systems, MDA, Canada}
% \member{Senior Member,~IEEE}
\author{ST\'EPHANE MARTEL}
% \member{Fellow,~IEEE}
\affil{Satellite Systems, MDA, Canada} 
%% \author{FOURTH D. AUTHOR}
%% \affil{University of Colorado, Colorado, USA}
\receiveddate{This work has been supported by the National
Research Council Canada’s (NRC) High Throughput Secure Networks program within the Optical Satellite Communications Consortium Canada (OSC) framework, by MDA, and by Mitacs. The authors would like to thank Sylvain Raymond and Colin Bellinger
from the National Research Council Canada (NRC) and Sameera Siddiqui
from the Defence R\&D Canada (DRDC).}
% \receiveddate{Manuscript received XXXXX 00, 0000; revised XXXXX 00, 0000; accepted XXXXX 00, 0000.\\
% This paragraph of the first footnote will contain the date on which you submitted your paper for review, which is populated by IEEE. It is IEEE style to display support information, including sponsor and financial support acknowledgment, here and not in an acknowledgment section at the end of the article. For example, ``This work was supported in part by the U.S. Department of Commerce under Grant BS123456.'' }
%% \accepteddate{XXXXX XX XXXX}
%% \publisheddate{XXXXX XX XXXX}

% \corresp{The name of the corresponding author appears after the financial information, e.g. {\itshape (Corresponding author: M. Smith)}. Here you may also indicate if authors contributed equally or if there are co-first authors.}

% \authoraddress{The next few paragraphs should contain the authors' current affiliations, including current address and e-mail. For example, First A. Author is with the National Institute of Standards and Technology, Boulder, CO 80305 USA 
% (e-mail: \href{mailto:author@boulder.nist.gov}{author@boulder.nist.gov}). Second B. Author, Jr., was with Rice University, Houston, TX 77005 USA. He is now with the Department of Physics, Colorado State University, Fort Collins, CO 80523 USA (e-mail: \href{mailto:author@lamar.colostate.edu}{author@lamar.colostate.edu}). Third C. Author is with the Electrical Engineering Department, University of Colorado, Boulder, CO 80309 USA, on leave from the National Research Institute for Metals, Tsukuba 305-0047, Japan 
% (e-mail: \href{mailto:author@nrim.go.jp}{author@nrim.go.jp}).}

\authoraddress{D. Bhattacharjee, P. G. Madoery, A. U. Chaudhry, H. Yanikomeroglu, and G. Karabulut Kurt are with Non-Terrestrial Networks (NTN) Lab, Department of Systems and Computer Engineering, Carleton University, Canada, e-mail: \{dhirajbhattacharjee, pablomadoery, auhchaud, halim\}@sce.carleton.ca. G. Karabulut Kurt is also with the Poly-Grames Research Center, Department of Electrical Engineering,  Polytechnique Montr\'eal, Canada, e-mail: gunes.kurt@polymtl.ca. P. Hu is with the Digital Technologies Research Center, National Research Council Canada, Canada, email: Peng.Hu@nrc-cnrc.gc.ca. K. Ahmed and S. Martel are with Satellite Systems, MDA, Canada, email:\{khaled.ahmed, stephane.martel\}@mda.space.}

% \editor{Mentions of supplemental materials and animal/human rights statements can be included here.}
% \supplementary{Color versions of one or more of the figures in this article are available online at \href{http://ieeexplore.ieee.org}{http://ieeexplore.ieee.org}.}

% \markboth{AUTHOR ET AL.}{SHORT ARTICLE TITLE}
\maketitle

\begin{abstract}Low Earth orbit (LEO) satellite mega-constellations are beginning to include laser inter-satellite links (LISLs) to extend the Internet to the most remote locations on Earth. Since the process of establishing these links incurs a setup delay on the order of seconds, a static network topology is generally established well in advance, which is then used for the routing calculations. However, this involves keeping links active even when they are not being used to forward traffic, leading to a poor energy efficiency. Motivated by technological advances that are gradually decreasing the LISL setup delays, we foresee scenarios in which it will be possible to compute routes and establish dynamic LISLs on demand. This will require considering setup delays as penalties that will affect the end-to-end latency. In this paper, we present a non-linear optimization model that considers these penalties in the cost function and propose three heuristic algorithms that solve the problem in a tractable way. The algorithms establish different trade-offs in terms of performance and computational complexity. We extensively analyze metrics including average latency, route change rate, outage probability, and jitter, in Starlink’s Phase I version 2 constellation. The results show the benefit of adaptive routing schemes according to the link setup delay. In particular, more complex schemes are able to decrease the average end-to-end latency, in exchange for an increase in the execution time.  On the other hand, depending on the maximum values of tolerated latency, it is possible to use less computationally complex schemes which will be more scalable for the satellite mega-constellations of the future.
\end{abstract}


\section{INTRODUCTION}
\IEEEPARstart{W}{hile} the standardization of 6G is still an ongoing process, the IMT-2030 draft documents from the International Telecommunication Union (ITU) offer a glimpse into the key differentiators from 3G, 4G, and 5G. In addition to targeting an order of magnitude improvement in data rate and latency, 6G aims to introduce a groundbreaking objective of providing inclusive connectivity (bridging the digital divide), ubiquitous coverage (anytime, anywhere), reliability, and sustainability. Achieving these ambitious goals affordably necessitates a pivotal consideration: the seamless integration of terrestrial and non-terrestrial networks (NTN) \cite{9982444}. This imperative, coupled with the advancements in launcher and satellite technologies, has revitalized the interest in deploying mega-constellations of low-Earth orbit (LEO) satellites.  Interconnected through inter-satellite links (ISLs), they can form satellite mesh networks capable of providing connectivity to areas where terrestrial infrastructure is not cost-effective. \db{Moreover, laser inter-satellite links (LISLs) are gradually replacing radio frequency (RF) ISLs due to their numerous advantages, including higher link capacity, reduced antenna size, higher directivity, lower power consumption, and diminished susceptibility to interception and interference \cite{freespaceoptics}. An RF ISL can achieve a capacity of around 2.5 gigabits per second (Gbps) \cite{freespaceoptics,2.5gbpsRFISL}, whereas there exists LISLs technology carrying 10 Gbps of capacity \cite{freespaceoptics}, \cite{telesat}. In addition, recent studies show that the link capacity of LISL will increase rapidly towards hundreds of Gbps \cite{Patnaik2012IntersatelliteOW,AroraGoyal} and even terabits per second (Tbps) range \cite{1Tbps} with technological advancements. This will reduce the queuing delays significantly for LISL based network and thus perform better than an RF ISL based network in terms of latency and throughput.}

However, given the relative mobility of LEO satellites with respect to each other, and with respect to the ground stations, numerous challenges must be addressed \cite{9210567}. These include topology planning and establishment, and the computation, management, and utilization of time-varying routes to forward traffic. The present paper delves into the existing trade-offs when establishing LISLs to form on-demand routes capable of delivering the required quality of service (QoS) to end-users. Special importance is given to the incorporation and modeling of link setup delays. \db{Due to highly directional laser beams, establishing a new LISL takes a few seconds \cite{carrizo2020optical}. On the contrary, it could take less than 1 ms for an RF ISL \cite{huang2018cascade}. Thus, dynamically establishing an ISL has a major impact on latency for LISL based satellite networks rather than an RF based one.}

\begin{figure}[!t]
    \centering
    \includegraphics[width=0.45\textwidth]{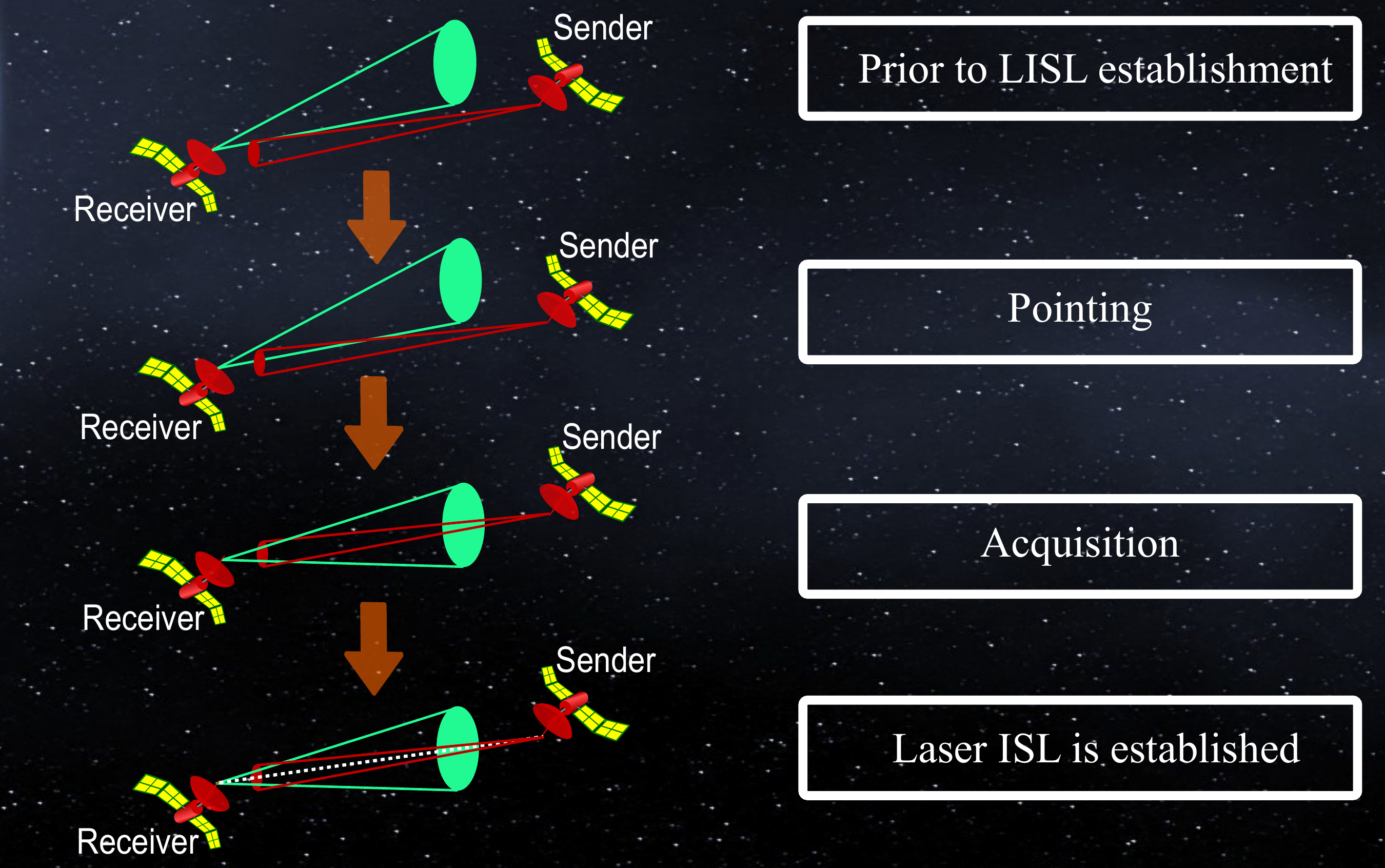}
    \caption{Pointing, acquisition, and tracking processes carried out in the establishment of a LISL. Sender satellite initiates the link establishment process by pointing its laser beam (shown by red color) towards receiver satellite. Also, receiver satellite aligns its laser beam (shown by green color) towards arriving beam. After accurate pointing and acquisition, both of the laser beams are aligned and the LISL is established.}
    \label{LISLestablishment}
\end{figure}

When two satellites are required to establish a LISL with each other, they follow a step-by-step procedure governed by the pointing, acquisition, and tracking (PAT) system \cite{kaushal2017free} as shown in Fig. \ref{LISLestablishment}. First, the sender satellite points its laser beam towards the receiver satellite and slowly scans the uncertainty area. Next, the receiver satellite aligns its laser beam towards the arriving beam and performs acquisition. Lastly, sender and receiver satellite continue pointing and acquisition process as the communication goes on and they track each others' laser beams \cite{PAT}. As a result of this procedure, the establishment of a new laser link requires a certain amount of time, hereby termed as LISL setup delay \cite{dhirajwcnc}. These delays represent a penalty that especially affects routes calculated on demand in energy-efficient routing schemes. By tackling these complexities, this paper aims to pave the way for a more seamless and energy-efficient integration of LEO mega-constellations, enabling global connectivity and ushering in the era of 6G wireless communication.
\subsection{Background and Related Work}

Although the first laser communication link between satellites was successfully demonstrated in 2001 by the European Space Agency \cite{FURCH2002223}, it was not until recently that such links began to be used in mega-constellations of satellites.  
Works such as \cite{freespaceoptics}, \cite{dhirajwcnc}, and \cite{9758829} have proposed classifications of LISLs and the different topologies that can be formed. From the topological point of view, LISLs can be considered as permanent or temporary \cite{chaudhry2022temporary}. While permanent links can be maintained indefinitely, temporary links may stop being feasible due to constraints such as distance or skew rate.
From a demand-based perspective, LISLs fall into two categories: static and dynamic \cite{dhirajwcnc}. A static LISL remains operational continuously, even during periods of inactivity (i.e., no traffic is sent through the link) by transmitting just enough power to maintain the link up for tracking purposes which is lower than the power required for communications. In contrast, a dynamic LISL is established in response to varying levels of traffic demand. In addition to these classifications, \cite{9758829} divides topologies into full grid-mesh, non-full grid-mesh, and non-grid-mesh. In the full grid-mesh topology, each satellite is connected to two neighboring satellites in the same orbital plane, and two neighboring satellites in adjacent orbital planes. In the non-full grid mesh topology, some of the previous links are disabled, in order to save energy, while maintaining end-to-end reachability. Finally, the non-grid-mesh topology is the most flexible scheme, but also the most challenging, since it implies a greater degree of freedom in establishing links between satellites in different orbital planes and with different relative velocities. 

Different studies have used these categories of LISLs to analyze topological and network metrics. Among the first papers, \cite{delaynotoptional} analyzes the use of LISLs in the Starlink constellation, assuming a full grid-mesh topology. This work performs an evaluation of the multiple routes that exist between arbitrary origins and  destinations, and analyzes the end-to-end latency considering the propagation time component. Complementing this work, \cite{9726866,10056385} evaluate and design crossover function with various scenarios in which lower latency can be obtained as compared to the use of terrestrial fiber networks.

In addition, \cite{10.1145/3359989.3365407} focuses on the problem of designing the topology, by means of link selection patterns, called motifs. It proposes schemes that have a reduced computational complexity and shows the trade-offs of the solutions in terms of latency and number of hops. Along the same lines, \cite{9758829} evaluates strategies to eliminate LISLs in order to improve the average bandwidth utilization, incurring a slight latency increase as a penalty. ISLs with higher length consume more power as the signal has to travel more and the received signal strength must be higher than a threshold to avoid outage. On the contrary, an end-to-end route will have lesser intermediate hops with longer ISLs which in turn will reduce the end-to-end delays. This latency versus power consumption trade-off in optical satellite networks is investigated in \cite{10354376,10287103}.

Furthermore, \cite{9914716} explores the performance and computational complexity of different on-demand routing algorithms that consider not only the propagation delay, but also the number of hops, which incur additional delays due to packet processing and queuing times. Among the contributions, a technique is proposed to reduce the search space and, therefore, the computational complexity. 

Finally, other works, such as \cite{9454162,9621840,8444381}, have begun to analyze the effect of traffic congestion and queuing delays on end-to-end latency, noting that these delays may contribute even more than the propagation delay. In \cite{9625325}, the authors propose a queuing delay and available bandwidth aware decentralized routing algorithm through reinforcement learning based training process. In addition, there could be heterogeneous applications which require different QoS and priority. \db{This necessitates treating the packets differently, and hence the routing algorithms need to adapt to it.} In \cite{9625013}, the authors present a QoS aware routing orchestration using \db{software-defined} networking based centralized approach. The algorithm considers QoS requirements of a flow such as delay, bandwidth, packet loss rate and jitter. Meeting end-to-end QoS requirement has two significant parts: handling the end-to-end decisions (routing) and handling the packets intelligently within a satellite node (queue management). Along with handling end-to-end routes according to type of a packet, the authors of \cite{9441632} also explore the merit of priority and weighted round robin scheduling to differentiate different types of packets.

The large body of recent work in this area reflects the importance of designing efficient routing schemes capable of dealing with the continuously changing topology in satellite mega-constellations. Most of the described optimizations only consider the propagation delay, and more recently, queuing and processing delays have been included in the cost function. However, to the best of the authors' knowledge, there has been no previous work that takes into account the link setup delay as an important component contributing to latency in routing schemes. This turns out to be the main motivation for the present paper.
\subsection{Motivation}
In general, setup delays have been overlooked due to the fact that the time required to establish LISLs is currently in the order of seconds \cite{carrizo2020optical}. Therefore, performing on-demand routing, together with \textit{just-in-time}\footnote{just-in-time is an inventory management method in which goods are received from suppliers only as they are needed.} topology establishment, would directly transfer these unacceptable delays to the end-to-end latency. Presently, Mynaric's laser terminals, known as CONDOR require 30 seconds to establish an LISL between two satellites during their initial interaction. Following the exchange of orbital parameters between the satellites, the subsequent establishment of a LISL takes around 2 seconds \cite{carrizo2020optical}. Meanwhile, both Tesat \cite{tesat} and General Atomics \cite{ga} have developed laser terminals intended for LEO constellations, featuring LISL setup delays in the range of tens of seconds. To avoid the impact of these delays on the QoS, most schemes assume that both topology establishment and route calculation are performed well in advance and are pre-configured in the satellites so that the links and routes are available when required.

Nevertheless, these approaches offer a low energy efficiency as they establish more links than necessary (and for longer), and compute an excessive number of routes. This results in a communication overhead, on-board satellite memory usage, and computational burden. We argue that the setup delays will decrease with the advancement of LISL technology in next-generation (NG) and next-after-next-generation (NNG) free-space optical satellite networks (FSOSNs) \cite{chaudhry2022temporary} by sophisticated scanning strategies for target acquisition, fast automatic calibration technologies, and optimized PAT operations \cite{IEEEaccess}, among others. With a reduced setup delay value, it will become increasingly feasible to compute and use on-demand routes to forward traffic. While this will allow the development of more energy efficient schemes, it will bring with it the challenge of considering the setup delay as part of the latency cost that needs to be minimized. 

Hence, the LISL setup delay emerges as a crucial factor in the formulation of on-demand routing algorithms, as it directly contributes to the overall end-to-end latency. The importance of the link setup delay was recognized in \cite{chaudhry2022temporary}, \cite{10.1145/3359989.3365407}, and \cite{impactonlatency} and a first study analyzing the implications was performed in \cite{dhirajwcnc}. The present work extends \cite{dhirajwcnc} significantly by designing and evaluating novel routing algorithms that provide different operating points in terms of performance and computational complexity. We also study how these solutions will behave in future scenarios in which the setup delay is expected to decrease from the order of seconds to the order of milliseconds.

\subsection{Contributions}
The main contributions of this work are as follows:

\begin{itemize}
    \item We formulate an optimization model for the on-demand routing problem in scenarios with dynamic LISLs, in non-grid mesh topologies, that introduce a link setup delay penalty on the end-to-end latency.
    \item We propose three heuristic algorithms that solve the optimization problem with varying performance and time complexity. 
    \item We analyze and compare the performance and time complexity of the three designed algorithms with a baseline algorithm in Phase I version 2 of the Starlink constellation. The analysis encompasses several performance metrics, including average end-to-end latency, latency distribution, route change rate, time complexity, outage probability, and jitter.
    \item We introduce emerging research directions that suggest expanding upon this study by developing routing schemes that address the anticipated requirements of forthcoming LEO satellite networks.
\end{itemize}

\vspace{0.4cm}

The remainder of the paper is organized as follows. Section~\ref{sec:network_arq} provides a description of the network architecture along with the variables and definitions used. Then, in Section \ref{sec:problem_form} these definitions are included in the formulation of the on-demand routing optimization problem. Subsequently, Section~\ref{sec:heuristic_algs} provides the design of three heuristic algorithms that are then evaluated and compared in Section \ref{sec:results}. Finally, the paper is concluded in Section \ref{sec:conclusion} by summarizing the main findings and motivating future work. 

\begin{figure}
    \centering
    \begin{subfigure}[b]{0.45\textwidth}
    \includegraphics[width=\textwidth]{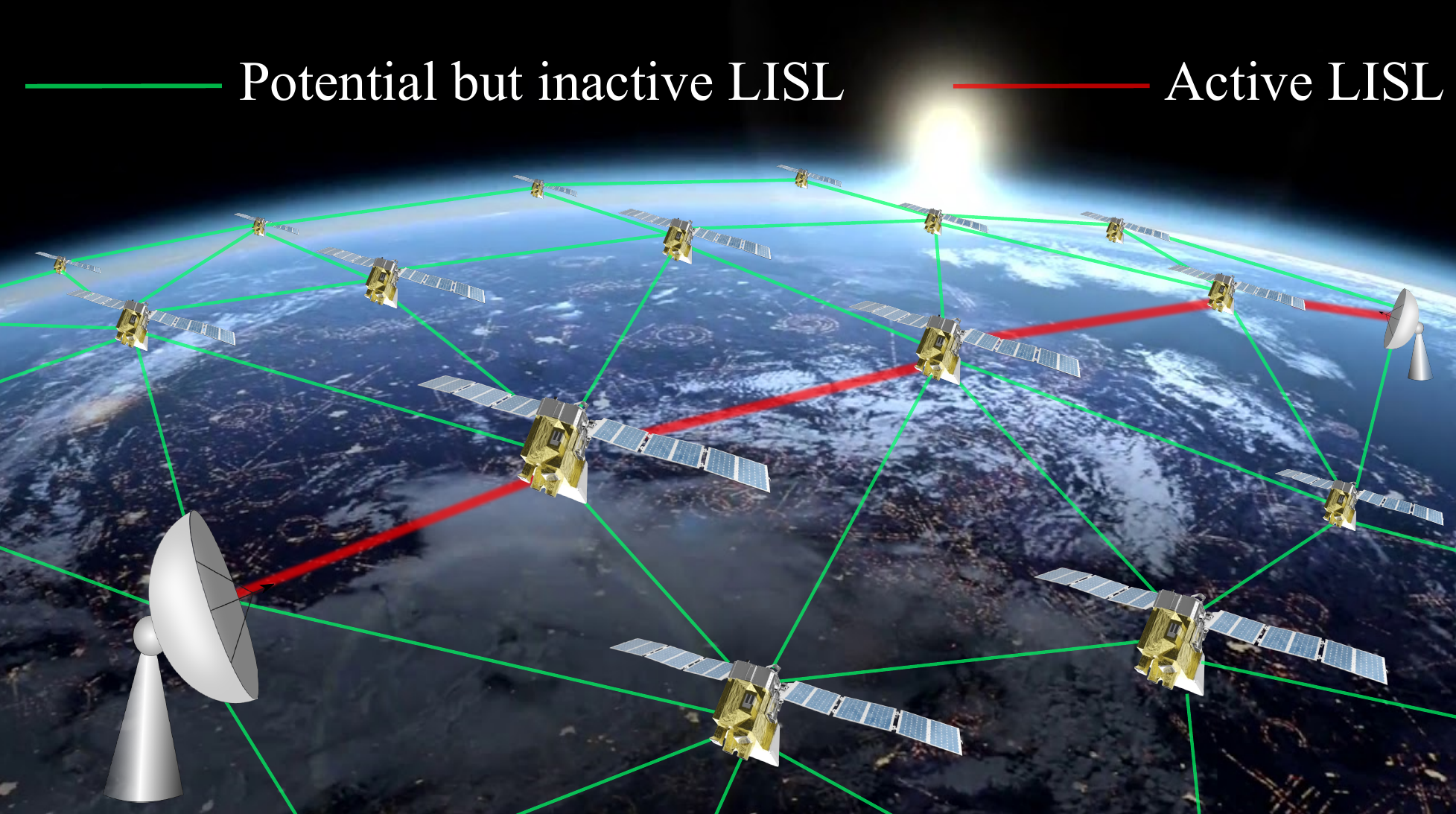}
    \end{subfigure}
    \caption{An exemplary network architecture where a source ground station communicates with a destination ground station. To establish an end-to-end route, only a subset (red lines) of all potential LISLs (green lines) is established.}
    
    \label{networkarchitecture}
\end{figure}

\section{NETWORK ARCHITECTURE}
\label{sec:network_arq}

We consider a network of LEO satellites with regenerative payload, and source and destination ground stations (GS) as ingress and egress point to the core satellite network as shown in Fig. \ref{networkarchitecture}. Each satellite is capable of establishing a LISL link with another satellite only when the distance between the two satellites is equal to or less than the LISL range. As a consequence, a flexible non-grid mesh topology is considered, where a satellite may establish connections with a varying number of neighbors. Similarly, a ground station can only establish an optical link with a satellite only if the distance between the satellite and the GS is equal to or less than the GS range. To capture the dynamic movement of LEO satellites, we consider a virtual topology model similar to those presented in \cite{s22124552, 634801, KORCAK20091497}. Through this approach, we capture the network's topology by taking $N$ snapshots, effectively dividing the simulation time into $N$ discrete time slots, each of which is denoted by the time index $i$. In each snapshot, the topology is assumed to be fixed. The cost of each edge consists of two components, propagation and node delays. Propagation delay is the length of the edge\footnote{The length of an edge is defined by the distance between two satellites associated with that edge.} divided by the speed of light in vacuum, and node delay is the sum of processing, transmission, and queuing delay. Processing delay is defined as the time taken by a node to process a packet to perform tasks such as reading packet header, finding the action from routing table, etc. Transmission delay can be calculated as the packet size divided by data rate of a link. Queuing delay is the time a packet is stored in the buffer before it gets the opportunity to be transmitted. Although when considering traffic, queuing delay will vary with time in different nodes, we assume a negligible queuing delay in order to explore the effect of LISL setup delay on routing decisions. A complete route, denoted by $r$, from a source GS to a destination GS is an ordered list of edges. The first edge of the list has the source GS as the source vertex, and the last edge has the destination GS as the target vertex. For all the other edges, the source vertex of the $z^{th}$ edge is the same as the target vertex of the $(z-1)^{th}$ edge. We also denote $\mathcal{R}$ as a set of all observable routes for a given source-destination pair during $N$ time slots. As the LEO satellites are mobile, the end-to-end delay of a route $\emph{r}$ changes with time and it is denoted as $\delta_{\emph{r}}^{[i]}$ at the $i^{th}$ time slot. $\delta_{\emph{r}}^{[i]}$ is the sum of the cost of all the edges associated with the route $r$.
 
In Fig. \ref{networkarchitecture}, a snapshot of the network is exemplified. Fig. \ref{networkarchitecture} shows if there is a traffic demand between the source and the destination GS, to establish an end-to-end route, only a few ISLs are established and rest of the ISLs are kept inactive. The ISLs in red colour represent the active edges whereas the green ISLs are inactive edges. Furthermore, we define a set of active LISLs as the edges which are active at time slot $i$ and we denote this as $\mathcal{A}^{[i]}$. For a technical clarification regarding static and dynamic LISLs, let us denote the set of all edges at time slot $i$ as $\mathcal{E}^{[i]}$. In the static LISL framework, all edges remain active at all times, meaning $\mathcal{A}^{[i]}=\mathcal{E}^{[i]}$. Conversely, within the dynamic LISL paradigm, only a specific subset of edges remain active, that is, $\mathcal{A}^{[i]}\subset\mathcal{E}^{[i]}$. Considering a scenario with one source-destination GS pair connection in the network, only the edges associated with a particular route $r$ will be active, leading to $\mathcal{A}^{[i]}\subset\mathcal{R}$. 

For routing table calculation and distribution orchestration, we consider a \db{software-defined} networking (SDN) architecture where GSs and satellites work as switches, and logically centralized controllers are physically distributed across the Earth. Ground controllers are responsible for calculating and distributing optimal routes, along with link information ($\mathcal{A}^{[i]}$), so that nodes (GSs and satellites) associated with those routes, can establish the necessary links. \db{Using the SDN architecture approach, ground controllers are able to communicate with GSs and satellites in a multi-hop fashion.} Initially, when a traffic demand reaches the source GS, there is no installed route. Therefore, the source GS contacts with a ground controller which in turn calculates and distributes the route, along with the links that must be established accordingly. \db{The source GS may not be within the communication range of the ground controller, but it can communicate through multiple satellites to reach the controller}. \db{Although this initial configuration imposes a delay} on the communication in the very beginning, it can be considered as a configuration mode, which is beyond the scope of this paper. \db{After configuring, we assume that the ground controller has the information of the traffic demands for which a route must continue to be provided.} \db{Now, as the optimal route from the source GS to the destination GS changes over time, the controller needs to calculate and distribute a new route before the route is required. This means that by the time a new route is required, the route and the link information need to be present in the associated nodes. Thus, we assume that the time stamp difference between two consecutive routes is greater than the time required to calculate the new route by the controller, plus the time required to distribute the new route and link information. Then, when the end-to-end route changes, new LISLs are established}. In addition to this, and in order to minimize size, weight and power (SWaP), we assume that satellites do not have spare optical terminals to serve a given demand. Therefore, \db{changing routes implies incurring a LISL setup delay}, denoted as $\eta_s$. Note that $\eta_s$ will be included in the end-to-end latency only once even if multiple LISLs are established. This is because the information on the new LISLs, which are to be activated, is already present in the respective nodes. Therefore, new LISLs can be established independently and simultaneously at the same time. For the sake of simplicity, we assume the same LISL setup delay value to establish any new LISL irrespective of the relative velocity between the associated nodes. Among all the routes in $\mathcal{R}$, if route $r$ is active at time slot $i$, we define a binary variable $\alpha_r^{[i]}$ as 1; otherwise, it is set to 0, as illustrated below:
\begin{equation}
    \alpha_r^{[i]}= \begin{cases}
                1      & r=\mathcal{A}^{[i]}\\ 
               0       & r\neq\mathcal{A}^{[i]}.
               
           \end{cases}
\end{equation}

All the variables and their significance are listed in Table \ref{notaion} for a quick reference.
\begin{table}
\caption{Notation \& Variables Used.}
\centering
\label{notaion}
\begin{tabular}{|c| c|}
\hline
\textbf{Notation} & \textbf{Definition}\\
\hline\hline
$N$ & Number of time slots \\
\hline
$i$ & Time slot index \\ 
\hline
$r$ & List of edges of a route \\
\hline
$\mathcal{R}$ & Set of all possible routes \\
\hline
$\delta_r^{[i]}$ & End-to-end delay of a route $r$ at time slot $i$ \\
\hline
$\mathcal{A}^{[i]}$ & Set of active edges at time slot $i$ \\
\hline
$\mathcal{E}^{[i]}$ & Set of all edges at time slot $i$ \\
\hline
$\eta_s$ & LISL setup delay \\
\hline
$\alpha_r^{[i]}$ & Route activation indicator \\
\hline
$\mathbf{D}$ & Delay matrix \\
\hline
$\mathbf{S}$ & Route selection matrix \\
\hline
$K$ & Number of all routes \\
\hline
$\lambda$ & Route change rate \\
\hline
$\eta_{delay}$ & End-to-end delay component for $N$ time slots\\
\hline
$\eta_{penalty}$ & Penalty component for $N$ time slots\\
\hline
$\eta_{LE}$ & Total end-to-end delay for $N$ time slots\\
\hline
$\overline{\eta_{LE}}$ & Average end-to-end delay per time slot\\
\hline
\end{tabular}
\end{table}

\section{PROBLEM FORMULATION}
\label{sec:problem_form}

First, we define the delay matrix $\mathbf{D}$ and route selection matrix $\mathbf{S}$ whose elements are $\delta_{\emph{r}}^{[i]}$ and $\alpha_{\emph{r}}^{[i]}$, respectively. Both matrices have $N$ number of columns and $|\mathcal{R}|=K$ number of rows as shown below:

\begin{equation}\label{Dmatrix}
    \mathbf{D}=
    \begin{pmatrix}
\delta_1^{[1]} & \delta_1^{[2]} & \delta_1^{[3]} & . & . & \delta_1^{[N]}\\
\delta_2^{[1]} & \delta_2^{[2]} & \delta_2^{[3]} & . & . & \delta_2^{[N]}\\
.&.&.&.&.&.\\
.&.&.&.&.&.\\
\delta_K^{[1]} & \delta_K^{[2]} & \delta_K^{[3]} & . & . & \delta_K^{[N]}
\end{pmatrix},
\end{equation}

\begin{equation}\label{Smatrix}
    \mathbf{S}=
    \begin{pmatrix}
\alpha_1^{[1]} & \alpha_1^{[2]} & \alpha_1^{[3]} & . & . & \alpha_1^{[N]}\\
\alpha_2^{[1]} & \alpha_2^{[2]} & \alpha_2^{[3]} & . & . & \alpha_2^{[N]}\\
.&.&.&.&.&.\\
.&.&.&.&.&.\\
\alpha_K^{[1]} & \alpha_K^{[2]} & \alpha_K^{[3]} & . & . & \alpha_K^{[N]}
\end{pmatrix}.
\end{equation}
As $r$ denotes the list of edges of a particular route, in (\ref{Dmatrix}) and (\ref{Smatrix}), rather than using $r$ in the subscripts, we use an integer route ID assigned to $r$. Some of the elements in $\mathbf{D}$ could be infinity as a route $r$ may not exist for a particular $i$ due to the mobility of LEO satellites. For example, $\mathbf{D}$ and $\mathbf{S}$ for $N=4$ and $K=3$ where each element of $\mathbf{D}$ represents the latency in milliseconds, are shown below:

% \begin{center}\label{example}
% \begin{tabular}{p{4cm}p{4cm}}
%   \begin{equation*}
%     \mathbf{D}=
%     \begin{pmatrix}
% 26 & 27 & 28 & \infty \\
% 27&26&25&25\\
% \infty &28&27&26\\
% \end{pmatrix},
%   \end{equation*}
%   &
%   \begin{equation}
%    \mathbf{S}=
%     \begin{pmatrix}
% 1&1&0&0\\
% 0&0&1&0\\
% 0&0&0&1\\
% \end{pmatrix}.
%   \end{equation} \\
  
% \end{tabular}
% \end{center}

\begin{equation}\label{example}
    \mathbf{D}=
    \begin{pmatrix}
26 & 27 & 28 & \infty \\
27&26&25&25\\
\infty &28&27&26\\
\end{pmatrix},\; \mathbf{S}=
    \begin{pmatrix}
1&1&0&0\\
0&0&1&0\\
0&0&0&1\\
\end{pmatrix}.
\end{equation}

The end-to-end delay component for one source-destination GS pair, and for $N$ time slots, is denoted as $\eta_{delay}$. The $\eta_{delay}$ can be calculated as follows:

\begin{equation}\label{delaycomp}
    \eta_{delay}=\sum_{i=1}^{N}\sum_{r\in\mathcal{R}} \delta_r^{[i]}\alpha_r^{[i]}.
\end{equation}

The penalty component for a source-destination GS pair, resulting from the LISL setup delay experienced over $N$ time slots, is  denoted as $\eta_{penalty}$.
The $\eta_{penalty}$ can be determined as follows:

\begin{equation}\label{penaltycomp}
    \eta_{penalty}=\eta_s\sum_{i=1}^{N-1}(1-\sum_{r\in\mathcal{R}}\alpha_r^{[i]}\alpha_r^{[i+1]}).
\end{equation}

To explain (\ref{penaltycomp}), we use the example of (4) where we consider a total of 3 available routes in 4 time slots. In the 1st and 2nd time slot, route 1 is active; in 3rd time slot route 2, and in 4th time slot route 3 is active. When the active route is the same (i.e., from time slot 1 to 2), $\sum_{r\in\mathcal{R}}\alpha_r^{[i]}\alpha_r^{[i+1]}$ is 1 leading $\eta_{penalty}$ to zero. Similarly, when the active route changes (i.e., from time slot 2 to 3, and from 3 to 4), $\sum_{r\in\mathcal{R}}\alpha_r^{[i]}\alpha_r^{[i+1]}$ is 0 leading $\eta_{penalty}$ to $\eta_s$.

Now, from (\ref{delaycomp}) and (\ref{penaltycomp}), we define the total end-to-end delay for $N$ time slots, $\eta_{LE}$, as the sum of $\eta_{delay}$ and $\eta_{penalty}$ given below:
\begin{equation}\label{totaldelay}
    \eta_{LE}=\eta_{delay}+\eta_{penalty}.
\end{equation}
Dividing (\ref{totaldelay}) by $N$, we get the average end-to-end delay as given below:
\begin{equation}\label{avgdelayeqn}
    \overline{\eta_{LE}}=\overline{\eta_{delay}}+\frac{\eta_s}{N}\sum_{i=1}^{N-1}(1-\sum_{r\in\mathcal{R}}\alpha_r^{[i]}\alpha_r^{[i+1]}).
\end{equation}

From (\ref{penaltycomp}), we define route change rate, $\lambda$, as how frequently the end-to-end route is changing and it is represented in percentage as shown below:
\begin{equation}\label{pathchangerate}
    \lambda=\frac{1}{N}\sum_{i=1}^{N-1}(1-\sum_{r\in\mathcal{R}}\alpha_r^{[i]}\alpha_r^{[i+1]})\times100\%.
\end{equation}
Using (\ref{avgdelayeqn}) and (\ref{pathchangerate}), $\overline{\eta_{LE}}$ can be related to $\lambda$ as follows:

\begin{equation}\label{avgdelayandpathchangerate}
    \overline{\eta_{LE}}=\overline{\eta_{delay}}+\frac{\eta_s}{100}\lambda.
\end{equation}

With the objective of minimizing $\eta_{LE}$, from (\ref{totaldelay}), we formulate the following minimization problem:

\begin{mini!}|s|
    {\scaleto{\alpha_r^{[i]}}{10pt}}{ \sum_{i=1}^{N}\sum_{r\in\mathcal{R}} \delta_r^{[i]}\alpha_r^{[i]}+\eta_s\sum_{i=1}^{N-1}(1-\sum_{r\in\mathcal{R}}\alpha_r^{[i]}\alpha_r^{[i+1]})}{}{}\label{p1}    
  \addConstraint{\sum_{r\in\mathcal{R}}\alpha_r^{[i]}\;=\;1;\;\forall i=1,2,...,N}\label{p1c1}
    \addConstraint{\alpha_r^{[i]}\in (0,1);\;\forall\;r\in\mathcal{R},i=1,2,...,N.}\label{p1c2}
\end{mini!}

The minimization problem (\ref{p1}) finds the optimal $\alpha_r^{[i]}$ $\forall\;r,\;i$. (\ref{p1c1}) ensures there is only one active route at each time slot and (\ref{p1c2}) defines $\alpha_r^{[i]}$ as a binary variable. Due to the existence of the sum of products of two binary variables in (\ref{p1}), this problem is an integer non-linear programming (INLP) problem. Now, to convert it to an integer linear programming (ILP) problem, we substitute the product of two binary variables as a single binary variable, and the necessary constraints are as follows:

\begin{mini!}|s|
    {\scaleto{\alpha_r^{[i]},\beta_r^{[i]}}{10pt}}{ \sum_{i=1}^{N}\sum_{r\in\mathcal{R}} \delta_r^{[i]}\alpha_r^{[i]}+\eta_s\sum_{i=1}^{N-1}(1-\sum_{r\in\mathcal{R}}\beta_r^{[i]})}{}{}\label{p2}    
  \addConstraint{\sum_{r\in\mathcal{R}}\alpha_r^{[i]}\;=\;1;\;\forall i=1,2,...,N}\label{p2c1}
    \addConstraint{\alpha_r^{[i]}\in (0,1);\forall\;r\in\mathcal{R},i=1,2,...,N}\label{p2c2}
    \addConstraint{\beta_r^{[i]}\leq\alpha_r^{[i]};\;\forall\;r\in\mathcal{R},}\;i=1,2..., N-1\label{p2c3}
    \addConstraint{\beta_r^{[i]}\leq\alpha_r^{[i+1]};\;\forall\;r\in\mathcal{R},}\;i=1,2..., N-1\label{p2c4}    \addConstraint{\beta_r^{[i]}\geq\alpha_r^{[i]}+\alpha_r^{[i+1]}-1;\;\forall\;r\in\mathcal{R},\;}{}\label{p2c5}\nonumber  
 \breakObjective{\qquad \qquad \qquad \qquad \quad i=1,2...,N-1.}{}{} 
\end{mini!}

Now, (\ref{p2}) represents an ILP problem and ILP problems are proven to be NP-hard \cite{hartmanis1982computers},\cite{schrijver1998theory}. As the route calculation has to be done in a limited amount of time in order for a new route to be present in the respective nodes before it is required, exhaustive search of this NP-hard problem is not possible. Thus, to solve (\ref{p2}) with a tractable complexity, we propose three heuristic algorithms that are discussed in the next section.
\section{PROPOSED HEURISTIC ALGORITHMS}
\label{sec:heuristic_algs}
\subsection{General Principles} Before discussing the technical descriptions of the proposed algorithms, we first provide a general overview of all presented algorithms. Each algorithm has two orthogonal perspectives: how the route is selected, and how long the selected route is active. Route selection can be based on the instantaneous or average latencies of the routes. The route, once selected, can be kept active for as long as possible (i.e., multiple time slots) or route selection can be performed in each time slot. In our benchmark algorithm, we apply Dijkstra's shortest route (DSR) algorithm \cite{dijkstra1959note} in each time slot. As this algorithm finds the route based on instantaneous latencies and route selection is performed in every time slot, the benchmark algorithm is termed as instantaneous latency based slotted routing (ILSR) algorithm. \db{Considering a binary heap-based priority queue implementation} of DSR algorithm, the time complexity of ILSR can be written as $O(N(V+E)\log{}V)$, where $N$ is the number of timeslots, $V$ is the number of satellite nodes, and $E$ is the number of edges in the network.

Our first proposed algorithm termed as instantaneous latency based persistent routing (ILPR) also selects the shortest route using DSR based on instantaneous latencies, but keeps the selected route active as long as the route exists. In average latency based persistent routing (ALPR), route selection is based on the average latencies of the available routes, and the selected route is persistently used for as long as possible. Finally, in instantaneous stability and activeness based slotted routing (ISASR) algorithm, the route is selected based on the stability of the LISLs and whether the link is already active or not. In addition, ISASR finds the route in each time slot.

\subsection{Instantaneous Latency based Persistent Routing (ILPR)}
In ILSR, the end-to-end route is decided based only on the edge cost (propagation plus node delays); as a result, the end-to-end route changes whenever the shortest route changes. This significantly affects the end-to-end latency, particularly when the LISL setup delay is high. To tackle this, we design ILPR as an elementary algorithm to avoid unnecessary link establishment for a high $\eta_s$ value and to be persistent on the selected route from source to destination as long as the route exists. In general, ILPR searches for the shortest route from source to destination and keeps that route as active as long as possible. When the route breaks (i.e., when any of the edges in the route no longer exist in the new time slot), ILPR calculates the shortest route at that time slot and keeps that route as the active route. 

\begin{algorithm}[t]
\caption{ILPR}\label{PRA}
\textbf{Input:}  $N$, EdgeDelay\\
\textbf{Initialization: } ${\alpha_r^{[i]}}$ = 0, $\forall\;i,r$\\
\textbf{Output:} $\alpha_r^{[i]};\;\forall\;i,r$

\begin{algorithmic}[1]
\FOR{$i$=1 to $N$}
    
    \IF{i==1}
        % \STATE ShortestPath=DSR(EdgeDelay$[i]$)
        \STATE $\mathcal{A}^{[i]}$=DSR(EdgeDelay$[i]$)
    \ELSE
        \IF{$\mathcal{A}^{[i-1]} \subseteq$ EdgeDelay$[i]$}
            % \STATE ShortestPath=DSR(EdgeDelay$[i]$)
            \STATE $\mathcal{A}^{[i]}=\mathcal{A}^{[i-1]}$
        \ELSE    
            \STATE $\mathcal{A}^{[i]}$=DSR(EdgeDelay$[i]$)
        \ENDIF
    \ENDIF 
    \STATE ${\alpha_r^{[i]}}\Big|_{r=\mathcal{A}^{[i]}}$ =1
    
    % For the route $\mathcal{A}^{[i]}$, set $\alpha_r^{[i]}$=1 and set $\alpha_r^{[i]}$=0 for rest of the routes in $\mathcal{R}$
\ENDFOR

\end{algorithmic}
\end{algorithm}

We present the steps of ILPR in Algorithm \ref{PRA}. As inputs, we need the number of time slots we want to solve the problem, $N$, and all the edges with the corresponding delay information for all $N$ time slots as EdgeDelay. Also, we initialize zero to $\alpha_r^{[i]}, \forall\;i,r$. For each time slot $i$, EdgeDelay$[i]$ represents a key-value pair mapping with edge as the key and delay as the value. At the beginning (i.e., the $1^{st}$ time slot), we apply DSR on EdgeDelay$[i]$ to obtain the list of edges as the current active route $\mathcal{A}^{[i]}$ (lines 2 to 3). After the first time slot, we check whether the old active route $\mathcal{A}^{[i-1]}$ exists. \db{If $\mathcal{A}^{[i-1]}$ is a subset of EdgeDelay$[i]$, it means that all the edges of $\mathcal{A}^{[i-1]}$ exist in the current timeslot $i$}. Thus, we retain the old active route as the current active route (line 6). On the contrary, if any of the edges does not exist in current timeslot $i$, (i.e., the old active route no longer exists), we again apply DSR on EdgeDelay$[i]$ to calculate the current shortest route as the new active route $\mathcal{A}^{[i]}$ (line 8). Finally, we set $\alpha_r^{[i]}$=1 for the route $r=\mathcal{A}^{[i]}$ (line 11).

For time complexity analysis of ILPR, an outer loop iterates N times. Line 3 introduces a complexity of $O((V+E)\log{}V)$. In a worst-case scenario, the condition check in line 5 consistently fails, resulting in the repeated invocation of DSR $N$ times. The complexity of the condition check in line 5 depends on the number of edges in $\mathcal{A}^{[i-1]}$, expressed as $O (V)$ in worst case. And finally, the complexities of straightforward lines like 2 and 11 are constant, denoted as $O(1)$. Thus, the worst-case time complexity of ILPR can be expressed as $O(N((V+E)\log{}V+V+1))=O(N(V+E)\log{}V)$.
% \begin{algorithm}
% \caption{PRA}\label{PR}
% \textbf{Input:}  $N$, EdgeDelay$[i]$ $\forall i$\\
% \textbf{Initialization: } $\mathcal{A}^{[i]}$=\{$\phi$\}, $\forall\;i$\\
% \textbf{Output:} $\alpha_p^{[i]};\;\forall\;i,p$

% \begin{algorithmic}[1]
% \FOR{$i$=1 to $N$}
%     \STATE ShortestPath=\textbf{DSR}\{EdgeDelay$[i]\}$
%     \IF{i==1}
%         \STATE Concatenate ShortestPath to $\mathcal{A}$
%     \ELSE
%         \IF{Shortest Path$\neq\mathcal{A}^{[i-1]} \&$ all elements in $\mathcal{A}^{[i-1]}$ do not exist in EdgeDelay$[i]$}
%             \STATE Concatenate ShortestPath to $\mathcal{A}$
%         \ELSE    
%             \STATE $\mathcal{A}^{[i]}=\mathcal{A}^{[i-1]}$
%         \ENDIF
%     \ENDIF 
%     \STATE For the path $\mathcal{A}^{[i]}$, set $\alpha_p^{[i]}$=1 and set $\alpha_p^{[i]}$=0 for rest of the paths in $\mathcal{P}$
% \ENDFOR

% \end{algorithmic}
% \end{algorithm}

\subsection{Average Latency based Persistent Routing (ALPR)}
It is important to note that ILPR only handles a high value of $\eta_s$ by avoiding a change in route as much as possible. However, for NG (mid to late 2020s) and NNG (early to mid 2030s) satellite mega-constellations, when $\eta_s$ gradually decreases, a more appealing approach would be a routing algorithm that is adaptive to the value of $\eta_s$. In addition, selecting the shortest route as the end-to-end route by ILPR is not optimal. This is because the shortest route is based only on the delay of the edges in the network, without considering the penalty $\eta_s$. Motivated by these shortcomings of ILPR, we present an approach adaptive to $\eta_s$ in Algorithm \ref{ACAR} termed as ALPR. In ALPR, instead of selecting the instantaneous shortest route, we consider the long-term cost of different routes, select the average shortest route, and keep that route active for as long as it exists. The fundamental working principle of ALPR is as follows: (i) ALPR computes the average latency of different routes including $\eta_s$ for the time slots a route exists, (ii) ALPR selects the route with the least average latency and uses the route until it expires, and (iii) ALPR only considers disjoint routes in the search process to reduce the complexity.
\begin{table}[t]
\caption{Example of ALPR Principle.}
\centering
\label{ACARexample}
\begin{subtable}[c]{.5\textwidth}
\caption{List of instantaneous delays.}\label{ACARinsdelay}
\centering
\begin{tabular}{|c| c|}
\hline
\textbf{Route} & \textbf{End-to-end delay}\\
\hline\hline
Route 1 & \{26,26.5,26.8,27,27.2,27.4\}\\
\hline
Route 2 & \{26.5,26.6,27.2,27.6,27.8,28.1,28.3,28.4,28.7,28.9,29.1\}\\
\hline
Route 3 & \{26.6,26.9,27.5,27.8,28,28.1,28.4\}\\
\hline
Route 4 & \{27.1,27.2,27.4,27.9,28.2,28.4,28.7,28.9\}\\
\hline
\end{tabular}
\vspace{2mm}

\end{subtable}

% \quad%
\begin{subtable}[c]{.5\textwidth}
\centering
\caption{Average delays for different LISL setup delays.}\label{ACARavgdelay}
\begin{tabular}{|c| c| c|}
\hline
\textbf{Route} & $\bm{\eta_s=1}$ \textbf{ms} & $\bm{\eta_s=1000}$ \textbf{ms}\\
\hline\hline
Route 1 & 26.98 & 193.48\\
\hline
Route 2 & 28.02 & 118.84 \\
\hline
Route 3 & 27.76 & 170.47\\
\hline
Route 4 & 28.1 & 152.97\\
\hline
\end{tabular}  
\vspace{2mm}

\end{subtable}
\end{table}

To explain the working principle of ALPR, we present an example in Table \ref{ACARexample}. In Table \ref{ACARinsdelay}, we list the end-to-end delay values for all available routes (assuming 4). It is important to note that the number of delay values for the routes could be different because a route may expire with time. Now, the question is among these routes which one to select. To determine this, we calculate the average latency as the sum of all the end-to-end delays of a route plus $\eta_s$, and then divide by the number of time slots the route exists. The route with the lowest average latency is selected for the time duration in which the route exists. Table \ref{ACARavgdelay} shows the average delay of 4 routes with two different $\eta_s$ values. We note that when $\eta_s$ value is small, the route with the least instantaneous end-to-end delays (i.e., route 1) has the least average delay. This is because, when $\eta_s$ value is small, averaging leads to a higher priority for instantaneous end-to-end delay compared to $\eta_s$. Therefore, in this case, the route with the least instantaneous end-to-end delay is selected. On the contrary, when $\eta_s$ value is high, more priority is given to $\eta_s$ and the longest existing route will be selected (i.e., route 2). Ideally, we should consider all possible routes from the source to destination GS but in mega-constellations that is not a scalable approach. Keeping this in mind, we only consider disjoint routes from the source to destination GS, as disjoint routes will be limited in number. To calculate the disjoint routes, first the shortest route from source to destination GS is calculated. Then all the edges associated with that shortest route are removed from the network topology. Next, again a new shortest route is calculated from the updated network topology and this process is continued until no more end-to-end route exists.

In Algorithm \ref{ACAR}, the outline of ALPR is illustrated. ALPR takes one additional input $\eta_s$ along with that of ILPR. Initially, we calculate the number of disjoint routes from the source to destination GS as the minimum number of edges associated with the source and destination GS (line 2). From lines 3 to 6 in Algorithm \ref{ACAR}, we first calculate the shortest route by applying DSR to EdgeDelay$[i]$ (line 4). Next, we remove all the edges of that route from EdgeDelay[i] to find a new disjoint shortest route (line 5). This process is repeated for the number of disjoint routes to obtain all the disjoint routes. Next, from lines 7 to 9, for each route in \textbf{DisjointRoutes}, we calculate the average end-to-end latency including $\eta_s$ for the time slots in which the route exists. To calculate the average end-to-end latency including $\eta_s$ of a route $r$ at time slot $i$, we sum all the end-to-end latency values from the current time slot $i$ to the last time slot the route exists (denoted as $l$) along with the LISL setup delay $\eta_s$ and then divide this sum by the number of time slots the route exists from the current time slot. Mathematically, the average end-to-end latency including $\eta_s$ denoted by $\overline{\eta}_r^{[i]}$ is calculated as follows:

\begin{equation}\label{avglatencyADR}
    \overline{\eta}_r^{[i]}=\frac{1}{l-i+1}(\eta_s+\sum_{k=i}^l \delta_r^{[k]}).
\end{equation}

After we obtain the average end-to-end latencies of all the disjoint routes, we select the route with the least average end-to-end latency (line 10) and keep the route active from time slot $i$ to $l$ (line 12). Finally, we set the time slot index $i$ as $l$ because the end-to-end route is to be determined from time slot $l+1$ (line 15). 

Now we analyze the worst-case time complexity of ALPR. The execution of line 2 involves a time complexity of $O(E)$, as all the edges must be checked to calculate the number of disjoint routes. Lines 3 to 6 can be executed a maximum of $V-1$ times, considering the fact that in an undirected graph, the maximum number of edges a node could have is $V-1$. The execution of line 4 has a complexity of $O((V+E)\log{} V$). Since a route can have a maximum number of edges on the order of $V$, worst-case time complexity of line 5 is $O(V)$. Moving forward, line 8 iterates in order of $V$ in the worst-case scenario. In line 8, first the delay values of the edges associated with a route need to be accessed, with a worst-case complexity of $O(V)$. This operation needs to be repeated for the duration the route exists, which could be on the order of $N$ at maximum. Calculating the average of these N latency values contributes to a time complexity of $O(VN+N)$. Due to a maximum of $V$ number of disjoint routes, line 10 requires a complexity of $O(V)$. For the worst case scenario, the selected route exists for only one timeslot, making the rest of the algorithm $O(1)$ in complexity. However, the outermost for loop is executed $N$ times. Combining all of these components, we express the worst-case time complexity of ALPR as $O(N(E+V((V+E)\log{}V+V)+V(VN+N)+V+1))=O(NV(NV+(V+E)\log{}V))$. 

\begin{algorithm}[t]
\caption{ALPR}\label{ACAR}
\textbf{Input:}  $N,\;\eta_s$, EdgeDelay\\
\textbf{Initialization: } ${\alpha_r^{[i]}}$ = 0, $\forall\;i,r$\\
\textbf{Output:} $\alpha_r^{[i]};\;\forall\;i,r$
\begin{algorithmic}[1]
\FOR{$i=1$ to $N$}
    % \STATE \%to find out all the disjoint paths\%
    % \IF{$RM==1$}
        \STATE NumDisjointRoutes=min(Src.edges,Dst.edges)
        % Calculate the number of disjoint paths from source to destination GS
        % \STATE Initialize DisjointPathDetails as \{$\phi$\}
        
        \FOR{$j=1$ to NumDisjointRoutes}
            \STATE \textbf{DisjointRoutes}[j]=DSR(EdgeDelay$[i]$)
            % Find ShortestPath by applying DSR on EdgeDelay$[i]$ and add to DisjointPathDetails
            % \STATE Concatenate ShortestPath to DisjointPathDetails
            % \STATE Set DisjointPathDetails$[j]$.ExistanceIndicator=1
            \STATE Remove edges in \textbf{DisjointRoutes}[j] from EdgeDelay$[i]$
        \ENDFOR
        % \STATE Set $RM=0$
    % \ENDIF
    % \STATE \%to find out which path to select\%

    % \IF{$RM==0$}
        \FOR{$j=1$ to NumDisjointRoutes}
            \STATE Calculate average end-to-end latency of \\ \textbf{DisjointRoutes}[j]  using (\ref{avglatencyADR}) 
            % of the path including $\eta_s$ from timeslot $i$ to the last timeslot the path exists.
        \ENDFOR
        \STATE $j_{min}=\arg \min_{j}$\big\{\textbf{DisjointRoutes}[j].AvgLatency\big\}
        
        \FOR{$k=i$ to $l$} 
            \STATE $\mathcal{A}^{[k]}=$\textbf{DisjointRoutes}[$j_{min}$]
            \STATE ${\alpha_r^{[k]}}\Big|_{r=\mathcal{A}^{[k]}}$ =1
        \ENDFOR
        % Select the route having least average latency and set as active route from timeslot $i$ to the last timeslot the route exists.
        \STATE $i=l$
        % \IF{DisjointPathDetails$[j]$.ExistanceIndicator==0, $\forall j$}
        %     \FOR{$j=1$ to number of disjoint paths}
        %         \STATE DisjointPathDetails$[j]$.AverageDelay=\{sum of DisjointPathDetails$[j]$.Delay+$\eta_s$\}/length of DisjointPathDetails$[j]$.Delay
        %     \ENDFOR
        %     \STATE  Set min index=$\argminA_j$ DisjointPathDetails$[j]$.AverageDelay
        
        %     \FOR{$j=1$ to length of  DisjointPathDetails[min index].Delay}
        %      \STATE Concatenate DisjointPathDetails[min index].edges in $\mathcal{A}$
        %      \STATE Set $\alpha_p^{[i]}=1$ for the selected path and $\alpha_p^{[i]}=0$ for all the other paths in $\mathcal{P}$
        %     \ENDFOR
        %     \STATE Set $i=$ number of entries in $\mathcal{A}$
        %     \STATE Set RM=1
        % \ENDIF
    % \ENDIF
    % \STATE $i++$
\ENDFOR

\end{algorithmic}
\end{algorithm}

\subsection{Instantaneous Stability and Activeness based Slotted Routing (ISASR)}
In ALPR, we consider the disjoint shortest routes from the source to destination GS to reduce the complexity of the algorithm. However, there may exist a better route that shares edges between two disjoint routes, and due to considering only the disjoint shortest routes, we may lose a better route by not even considering it. This downside can be addressed by considering all possible routes, but the solution will not be scalable for mega-constellations. Thus, for a better and scalable solution, we design ISASR, where we do not consider all possible routes, but we modify the existing cost of edges by quantifying the stability and activeness of the individual edge and then apply DSR at each time slot. We also remove some edges from the network search space whose stability related cost is higher than a certain threshold value to reduce computational complexity. Modification of edge cost involves adding a weighted sum of the stability and activeness-related cost of an edge (denoted by cost$_{st}$ and cost$_{act}$, respectively) with the old cost. ISASR sets cost$_{st}$ of an edge proportional to $\eta_s$ and inversely proportional to the number of time slots in which the edge exists. This increases the modified cost of an edge when $\eta_s$ is high and/or the edge exists for fewer time slots. On the other hand, cost$_{act}$ is set as a positive value for an edge that is not active and set as zero for an active edge. Combining these costs with the old cost of an edge leads to the selection of more stable and already active edges when $\eta_s$ is high. Similarly, when $\eta_s$ is low, more unstable edges can be selected and more frequent route change is inevitable.

In the detailed step-by-step procedures of ISASR shown in Algorithm \ref{eCAR}, along with the required inputs mentioned in ALPR, additional inputs are required in ISASR. LinkDetails is a key-value paired map input where the key is an edge and the value is an array of time slot indices for which the edge exists. cost$_{thrsh}$ is a threshold value to compare the stability-based cost of an edge, cost$_{st}$ for deciding whether to keep or remove the edge from the search space, EdgeDelay. Furthermore, $\gamma$ is required as an input—a weight factor used to adjust the edge costs. As mentioned earlier, the old cost of an edge (referred to as cost$_{old}$) is modified to obtain cost$_{mod}$ as follows:
\begin{equation}\label{modifycost}
    \text{cost}_{mod}=\text{cost}_{old}+\gamma\{\text{cost}_{st}+\text{cost}_{act}\}.
\end{equation}

Initialization or updating of cost$_{st}$ follows (\ref{stabilitycost}), as shown below: 

\begin{equation}\label{stabilitycost}
    \text{cost}_{st}^{[i]}= \begin{cases}
                0      & l=N\\ 
               \frac{\eta_s}{l-f+1}       & i<f\; \text{and} \;l<N\\
               \frac{\eta_s}{l-i+1}       & f\leq i\leq l < N\\               
               \infty & l< i\leq N.\\
           \end{cases}
\end{equation}

In (\ref{stabilitycost}), denoting $f$ and $l$  as the first and the last time slot ($f\leq l$) an edge exists in the network, if $l=N$, signifying that a route will not break for the edge within $N$ timeslot and cost$_{st}$ is assigned as $0$. Conversely, if $l<N$ there are three possible cases: (i) current timeslot index $i$ is lesser than $f$, and cost$_{st}$ is set to $\eta_s$ divided by the number of timeslots the edge exists; (ii) $i$ is in between $f$ and $l$, and cost$_{st}$ is calculated as $\eta_s$ divided by the number of remaining timeslots the edge exists; and (iii) $i$ is greater than $l$, and cost$_{st}$ is set to $\infty$.

From line 1 to 4, cost$_{act}$ and cost$_{st}$ of all the edges in LinkDetails are initialized. Initially, as there is no active route, for each edge, cost$_{act}$ is set to $\eta_s$ (line 2) and cost$_{st}$ is initialized using (\ref{stabilitycost}) (line 3).

From line 6 to 11, search space, i.e., the size of EdgeDelay is reduced and costs of edges are updated accordingly. First, we check for edges between two satellites and have the stability cost (i.e., cost$_{st}$) higher than cost$_{thrsh}$. These edges are removed from EdgeDelay$[i]$ (lines 7 to 9). ISASR keeps all GS to satellite edges irrespective of their cost$_{st}$, as that could lead to lack of any connection between a GS to the satellites. Next, we modify the costs of the edges in EdgeDelay$[i]$ using (\ref{modifycost}) in line 10.

\begin{algorithm}[t]
\caption{ISASR}\label{eCAR}
\textbf{Input:}  $N$, $\eta_s$, LinkDetails, cost$_{thrsh}$, $\gamma$, EdgeDelay\\
% \textbf{Initialization: } LinkCostDetails, $\mathcal{A}^{[i]}$=\{$\phi$\}, $\forall\;i$\\
\textbf{Initialization: } ${\alpha_r^{[i]}}$ = 0, $\forall\;i,r$\\
\textbf{Output:} $\alpha_r^{[i]};\;\forall\;i,r$

\begin{algorithmic}[1]
\FOR{edge in LinkDetails}
    \STATE cost$_{act}=\eta_s$
    \STATE Initialize cost$_{st}$ using (\ref{stabilitycost})
    % \IF{edge exists $\forall i$}
    %     \STATE LinkCostDetails[edge].cost$_{st}=0$
    % \ELSE
    %     \STATE LinkCostDetails[edge].cost$_{st}=\eta_s$/number of time slots the edge exists
    % \ENDIF
\ENDFOR

\FOR{i=1 to $N$}
    \FOR{edge in EdgeDelay$[i]$}
        \IF{cost$_{st}\geq$cost$_{thrsh}$}
            \STATE Remove edge from EdgeDelay$[i]$
        \ENDIF
        \STATE Update cost of edge using (\ref{modifycost})
    \ENDFOR
    % \FOR{edge in RemoveLinks}
    %     \STATE Remove edge from EdgeDelay$[i]$
    % \ENDFOR
    % \FOR{edge in EdgeDelay$[i]$}

    %     % EdgeDelay$[i]$[edge][delay]=EdgeDelay$[i]$[edge][delay]+$\gamma$ \{LinkCostDetails[edge].cost$_1$+LinkCostDetails[edge].cost$_2$\}
    % \ENDFOR
    \STATE $\mathcal{A}^{[i]}$=DSR(EdgeDelay$[i]$)
    \STATE ${\alpha_r^{[i]}}\Big|_{r=\mathcal{A}^{[i]}}$ =1
    % Apply DSR on EdgeDelay$[i]$ and set the output path to 
    \STATE RouteBreakPoint=CalcRouteBreak($\mathcal{A}^{[i]}$, LinkDetails)
    % Find PathBreakPoint of $\mathcal{A}^{[i]}$
    \FOR{edge in $\mathcal{A}^{[i]}$}
        \IF{$i\neq$ RouteBreakPoint}
            \STATE cost$_{act}=0$
        \ELSE
            \STATE cost$_{act}=\eta_s$
        \ENDIF
    \ENDFOR
    \FOR{edge in LinkDetails}
        \STATE Update cost$_{st}$ using (\ref{stabilitycost})
     \ENDFOR
\ENDFOR
\end{algorithmic}
\end{algorithm}

After modifying the costs, we find $\mathcal{A}^{[i]}$ by applying DSR to EdgeDelay$[i]$ (line 12) and $\alpha_r^{[i]}$ accordingly (line 13). Now, we find the RouteBreakPoint of the selected route $\mathcal{A}^{[i]}$ which is the minimum time slot index of the last time slots that the edges in $\mathcal{A}^{[i]}$ exist (line 14). For example, let us assume that the last time slot indices that the active edges exist are $\{105,117,93,155,148\}$. Thus, RouteBreakPoint for this active route will be $93$ as this route will break after time slot $93$.

From line 15 to 21, cost$_{act}$ of only the edges of active route is updated as for the inactive edges, it will be unchanged. Until $i$ reaches to RouteBreakPoint, cost$_{act}$ of active edges are set to $0$ (line 17) so that active edges get priority of getting selected at the current time slot. When the time slot index reaches RouteBreakPoint, one or more edges of the active route expire, and the route breaks. At this point, as the inclusion of penalty is inevitable, cost$_{act}$ of all the edges should be the same, as there is no significance of giving priority to old active edges anymore. Hence, when the time slot reaches RouteBreakPoint, cost$_{act}$ of all the active edges is set as $\eta_s$ (line 19). After updating cost$_{act}$ of the active edges, we update the cost$_{st}$ of the edges using (\ref{stabilitycost}) in lines 22 to 24.

In a worst-case scenario, LinkDetails can reach a maximum size of $NE$, assuming all the links exist for only one time slot. Initialization of cost$_{act}$ and cost$_{st}$ each takes $O(1)$ in complexity, resulting in a total time complexity of executing lines 1 to 4 as $O(NE)$. Lines 6 to 11 have a time complexity of $O(E)$. In the worst case, no edge is removed from the search space, and EdgeDelay retains all $E$ entries contributing to a complexity of $O((V+E)\log{}V)$ in line 12. Moving forward, the number of edges in a route is, at most, in the order of $V$, as discussed previously. Therefore, line 14 has a time complexity of $O(V)$. Similarly, lines 15 to 21 and lines 22 to 24 include time complexities of $O(V)$ and $O(NE)$, respectively. Finally, considering that lines 5 to 25 are executed $N$ times, the worst-case time complexity of ISASR can be expressed as, $O(NE+N(E+(V+E)\log{}V+V+V+NE))=O(N(NE+(V+E)\log{}V))$.

\section{SIMULATION RESULTS}
\label{sec:results}

\subsection{Simulation Settings}
\begin{table}
\caption{Simulation Parameters.}
\centering
\label{simparam}
\begin{tabular}{|c| c|}
\hline
\textbf{Parameter} &\textbf{Value}\\
\hline\hline
Number of satellites & $1584$\\ 
\hline
Number of orbits & $24$ \\
\hline
Number of satellites per plane & $66$ \\
\hline
Orbit inclination & $53\degree$\\
\hline
Orbit altitude & $550$ km \\
\hline
Speed of a satellite & $7.6$ km/s \cite{chaudhry2021optical}\\
\hline
Satellite LISL range & $1500$ km\\
\hline
GS range & $1000$ km\\
\hline
Node delay & $1$ ms\\
\hline
$N$ & $600$ \\
\hline
Duration of a time slot & $1$ second\\
\hline
cost$_{thrsh}$ & 100\\

\hline
\end{tabular}
\end{table}

In this section, the simulation results for four contenders (ILSR, ILPR, ALPR, and ISASR) are discussed. We simulate Starlink's Phase I version 2 constellation of 1584 satellites using Ansys's Systems Tool Kit (STK) \cite{stk}. This constellation has 24 orbits and each of them has 66 satellites at 550 km altitude with 53$\degree$ of inclination angle \cite{spacex}. At 550 km of altitude, the satellite speed is calculated to be 7.6 km/s \cite{chaudhry2021optical}. We consider LISL range as 1500 km and GS range as 1000 km \cite{dhirajwcnc}. As 5016 km is the maximum LISL range for Starlink Phase I version 2 constellation calculated in \cite{9393372}, by keeping LISL range to 1500 km, we ensure that there is no Earth blockage to the ISLs. In this simulation, we use Walker Delta constellation with number of neighbours of a satellite varying with time due to satellites in motion. Considering very high data rate \db{(tens of gigabits per second)} LISLs, the transmission delay is assumed to be negligible. Also, as congestion is beyond the scope of this paper, queuing delay is not considered. In addition, the processing delay is set to be 1 ms \cite{1msdelay}, resulting in a total node delay in each satellite being 1 ms. All the simulation parameters are listed in Table \ref{simparam}. Using constellation parameters, first we generate the constellation in STK and by interfacing STK with Python, we extract constellation dataset e.g., vertices, edges, and length of edges for 600 time slots with each time slot being 1 second. Then the constellation dataset is used to perform the algorithms in Python. First, we show the average end-to-end latency performance of the four routing algorithms with different $\eta_s$ values for New York-London and New York-Hanoi inter-continental connections. Next, we show how these algorithms are adapted with different $\eta_s$ values in terms of balancing between the average delay component, $\overline{\eta_{delay}}$ and the route change rate, $\lambda$. In ISASR, the parameter $\gamma$ is kept proportional to $\eta_s$ and in the simulation, we consider $\gamma=\eta_s$. In this regard, we show a numerical proof and intuitive reasoning that $\gamma$ is to be proportional with $\eta_s$. We set cost$_{thrsh}$ to 100 unless specified otherwise by observing the histograms of cost$_{st}$ in such a way that no edges are removed to show the highest performance of ISASR. We also compare these algorithms in terms of time complexity and provide insights into performance versus complexity trade-offs. Finally, we present a comparative analysis of histograms of the instantaneous end-to-end latencies of these four algorithms along with outage probabilities and average jitter performance.
\subsection{Average End-to-End Latency}

\begin{figure}[!b]
    \centering
    \begin{subfigure}[b]{0.48\textwidth}
        \centering
        \includegraphics[width=\textwidth]{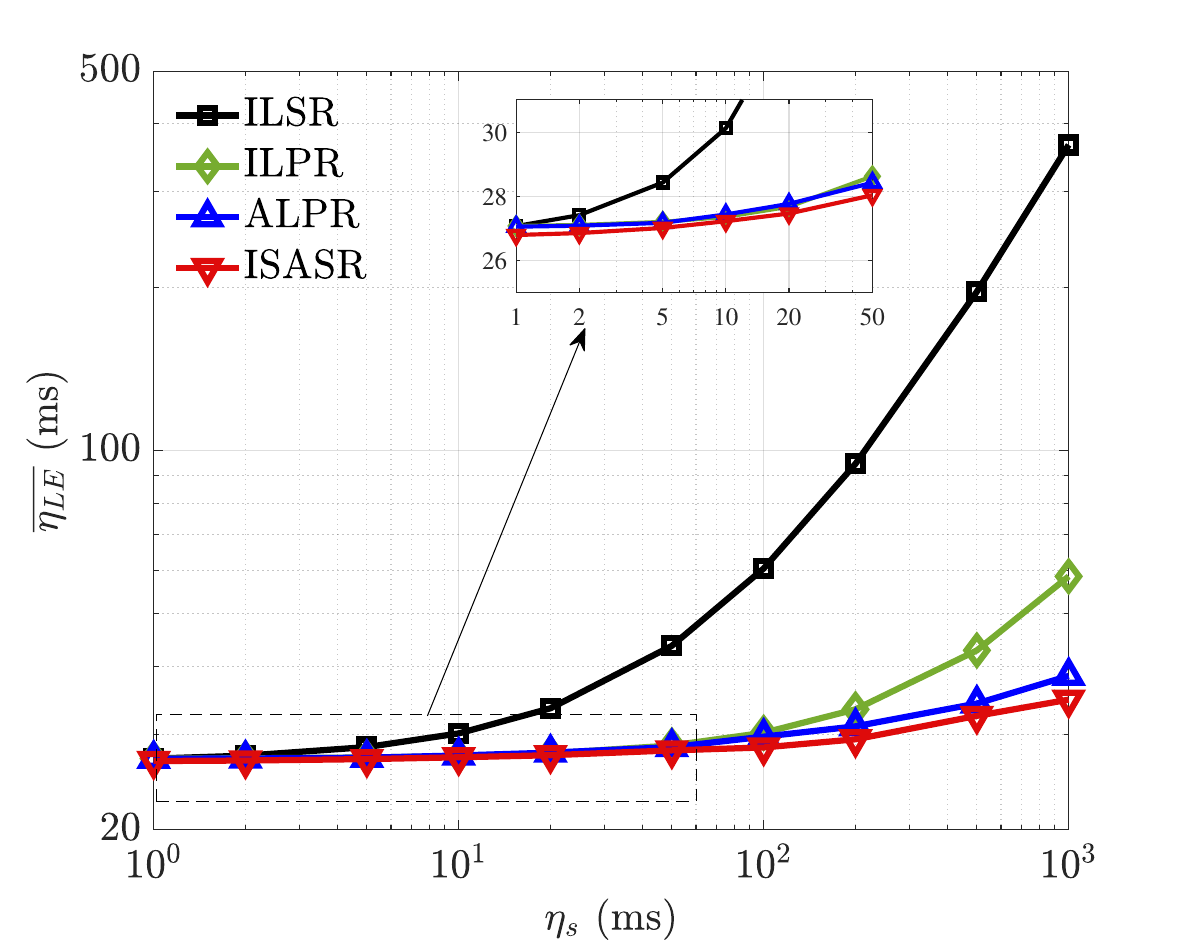}
        \caption{New York to London.}
        \label{avgdelaylondon}
    \end{subfigure}
    \quad
    \begin{subfigure}[b]{0.48\textwidth}
        \centering 
        \includegraphics[width=\textwidth]{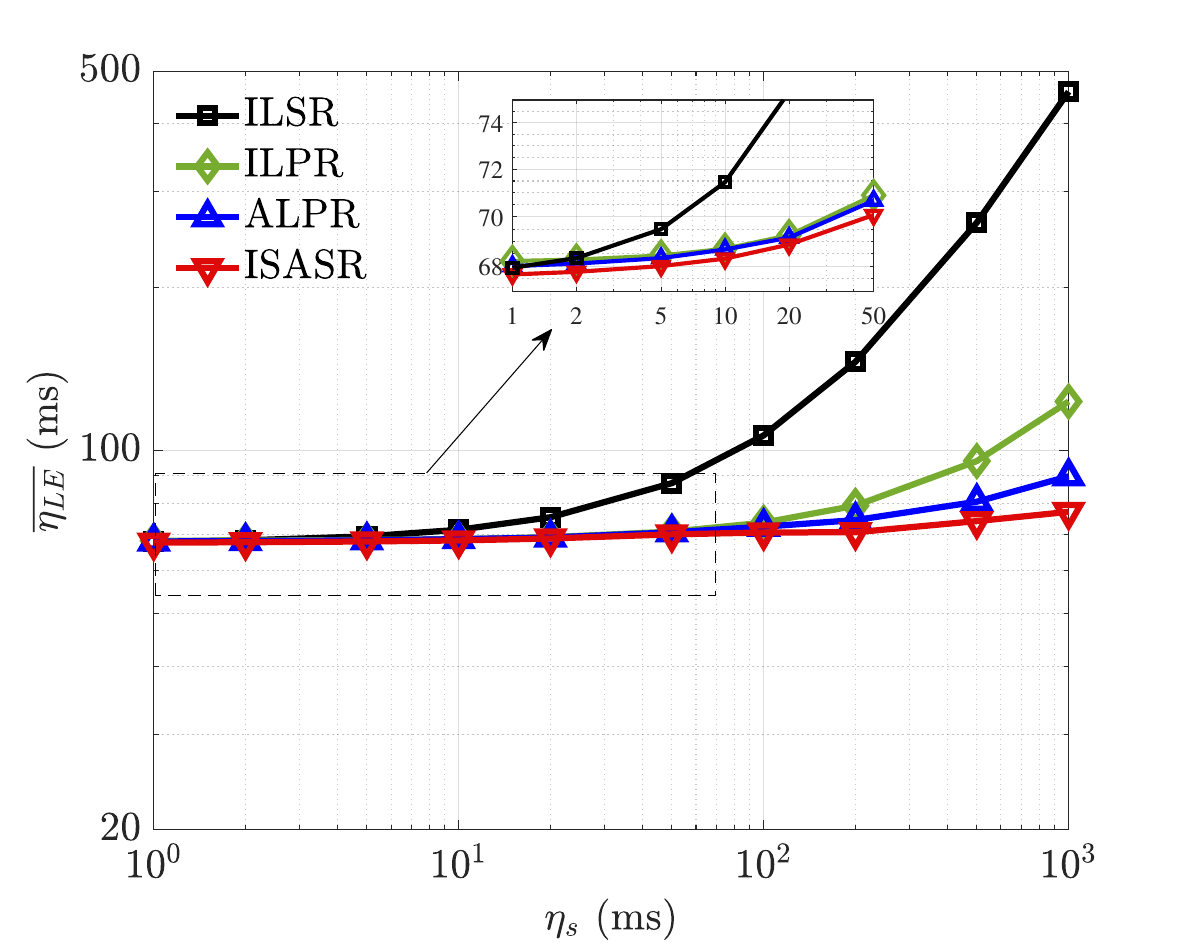}
        \caption{New York to Hanoi.}
        \label{avgdelayhanoi}
    \end{subfigure}
    % \vskip\baselineskip
    \caption{Plot of average end-to-end latency ($\overline{\eta_{LE}}$) vs LISL setup delay ($\eta_s$) for 2 inter-continental connections (New York-London and New York-Hanoi). The general trend shows that higher $\eta_s$ leads to a higher $\overline{\eta_{LE}}$ for all four algorithms (ILSR, ILPR, ALPR, and ISASR). For medium to high values of $\eta_s$, the $\overline{\eta_{LE}}$ is maximum for ILSR (depicted by the solid black line) and minimum for ISASR (represented by the solid red line). As shown in zoom-in squares, for very low values of $\eta_s$, ALPR (represented by the solid blue line) and ILPR (represented by the solid green line) have marginally higher values of $\overline{\eta_{LE}}$ compared to ILSR.}\label{avgdelay}
\end{figure}

Fig. \ref{avgdelay} shows average end-to-end latency $\overline{\eta_{LE}}$ variation with $\eta_s$ for New York -London (Fig. \ref{avgdelaylondon}) and New York-Hanoi (Fig. \ref{avgdelayhanoi}) inter-continental connection. Clearly, $\overline{\eta_{LE}}$ increases with $\eta_s$ as the penalty component increases with $\eta_s$. We can observe a very high value of $\overline{\eta_{LE}}$ for ILSR compared to the other three algorithms, unless $\eta_s$ is very small. This is because ILSR does not consider $\eta_s$ into account and changes the end-to-end route based only on the delay component, $\eta_{delay}$, which affects $\overline{\eta_{LE}}$ significantly for medium to high $\eta_s$ values\footnote{Low, medium, and high values of $\eta_s$ are in the range of 1 to 10 ms, 10 to 100 ms, and 100 to 1000 ms, respectively.}. In contrast, being in the selected old shortest route as long as possible by ILPR gives a significant improvement in $\overline{\eta_{LE}}$. As discussed earlier, ILSR and ILPR are not $\eta_s$ aware whereas ALPR and ISASR are. ILSR and ILPR always select the shortest route (current or old), whereas ALPR selects the route that is the best average-wise. Therefore, for high $\eta_s$, ALPR selects a more stable route that may not always be the shortest, and for low $\eta_s$, ALPR focuses more on the delay component rather than the penalty. Thus, we can see a clear improvement in ALPR as compared to ILSR and ILPR for medium to high $\eta_s$ for both inter-continental connections. In the lower range of $\eta_s$, as depicted in the zoomed-in portions, both ILPR and ALPR exhibit the same performance. At lower $\eta_s$ values, ALPR's approach to calculate the average delay for multiple disjoint routes assigns higher significance to instantaneous latency over the penalty, $\eta_s$. Consequently, this leads to the selection of the instantaneous shortest route and keeps the route active, effectively converging towards ILPR behavior as $\eta_s$ is reduced. An additional performance improvement is evident in ISASR across all $\eta_s$ values compared to ALPR. Interestingly, for $\eta_s=1$ ms, the performance of ILSR becomes slightly better than that of ILPR and ALPR. This is because of the tendency of ILPR and ALPR to stick with a chosen route even when better alternatives exist, and route changes incur only minor penalties. On the other hand, ISASR still outperforms ILSR for $\eta_s=1$ ms because of its adaptive cost modification and the flexibility to switch routes at any time slot. Comparing the end-to-end latencies in Fig. \ref{avgdelaylondon} and Fig. \ref{avgdelayhanoi} of a specific algorithm, the average latency of New York-Hanoi is higher than New York-London. This is straightforward because both the delay and penalty components of New York-Hanoi are greater compared to those of New York-London. As the delay component comprises propagation and node delays, a longer source-destination connection implies higher propagation delay along with increased node delay owing to a greater number of intermediate hops. Furthermore, a longer connection entails more edges along the end-to-end route, elevating the likelihood of route disruptions. Consequently, this contributes to an increased $\lambda$ value and subsequently a higher penalty component.

\begin{figure}[t]
    \centering
    \begin{subfigure}[b]{0.48\textwidth}
        \centering
        \includegraphics[width=\textwidth]{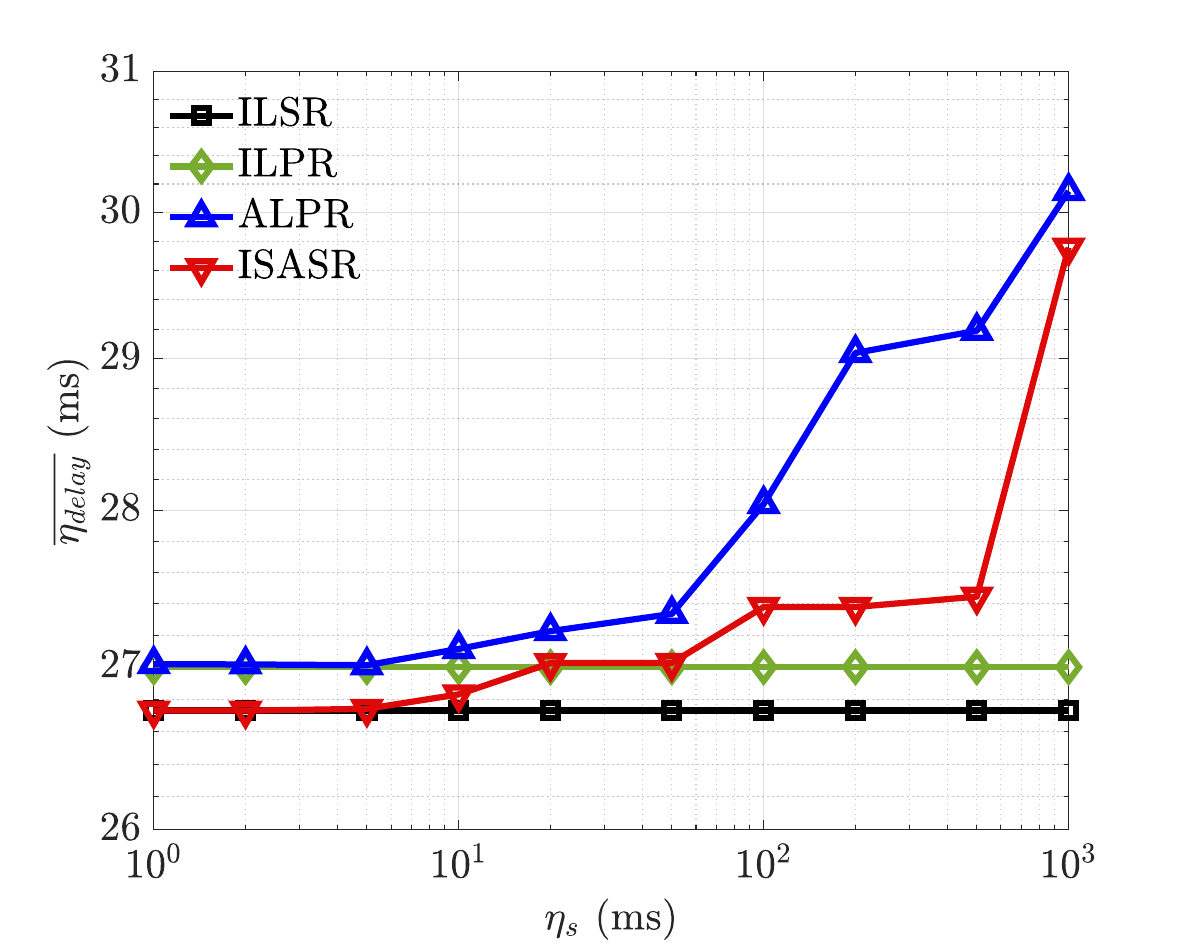}
        \caption{Average delay component.}
        \label{split1}
    \end{subfigure}
    \quad
    \begin{subfigure}[b]{0.48\textwidth}
        \centering 
        \includegraphics[width=\textwidth]{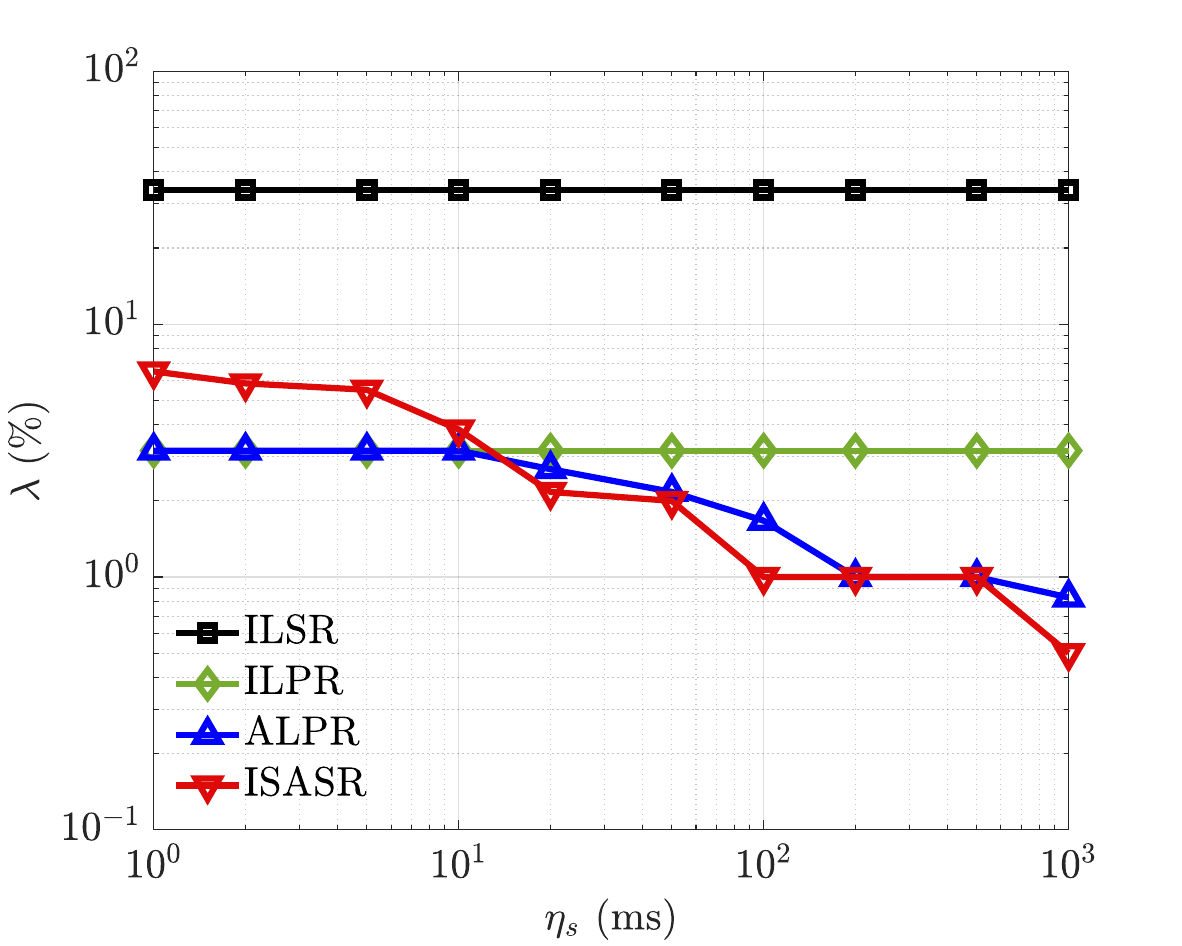}
        \caption{Route change rate.}
        \label{split2}
    \end{subfigure}
    % \vskip\baselineskip
    \caption{Plot of average delay component ($\overline{\eta_{delay}}$) and route change rate ($\lambda$) vs LISL setup delay ($\eta_s$) for New York-London inter-continental connection. Differently than ALPR (shown by solid blue line) and ISASR (shown by solid red line), ILSR (shown by solid black line) and ILPR (shown by solid green line) are not adaptable with $\eta_s$, thus having a constant $\overline{\eta_{delay}}$ and $\lambda$ value with respect to $\eta_s$. On the other hand,  for ALPR and ISASR, higher $\eta_s$ leads to a higher $\overline{\eta_{delay}}$ and lower $\lambda$. For med to high $\eta_s$ values, ALPR has maximum values of $\overline{\eta_{delay}}$ while ISASR has minimum values of $\lambda$.}\label{avgdelaysplit}
\end{figure}

Using (\ref{avgdelayandpathchangerate}), we present the breakdown of $\overline{\eta_{LE}}$ as shown in Fig. \ref{avgdelaysplit} of New York-London as the average delay component, $\overline{\eta_{delay}}$ and the route change rate, $\lambda$. This visualization is intended to offer enhanced clarity into the dynamics of the four algorithms. Starting with ILSR and ILPR algorithms, both $\overline{\eta_{delay}}$ and $\lambda$ exhibit no variations with respect to $\eta_s$ as they are not adaptive to $\eta_s$. Compared to ILSR, in ILPR, $\overline{\eta_{delay}}$ is slightly compromised to focus more on $\lambda$ for improved performance in $\overline{\eta_{LE}}$. However, this policy is effective only for high $\eta_s$ values, necessitating a proper balance between $\overline{\eta_{delay}}$ and $\lambda$ for any given $\eta_s$. Moving forward to ALPR and ISASR algorithms, we observe a greater degree of compromise on $\overline{\eta_{delay}}$ and more focus on $\lambda$ for high $\eta_s$ values. However, as $\eta_s$ is reduced, the emphasis shifts from $\lambda$ towards $\overline{\eta_{delay}}$ such that the combined effect of $\overline{\eta_{delay}}$ and $\lambda$ on $\overline{\eta_{LE}}$ is minimized. For lower values of $\eta_s$, both of the metrics $\overline{\eta_{delay}}$ and $\lambda$ of ILPR and ALPR tend to converge which essentially leads to the same average latency performance, as discussed earlier. 
\subsection{Selection of $\gamma$ in ISASR}

\begin{figure}[!t]
    \centering
    \includegraphics[width=0.48\textwidth]{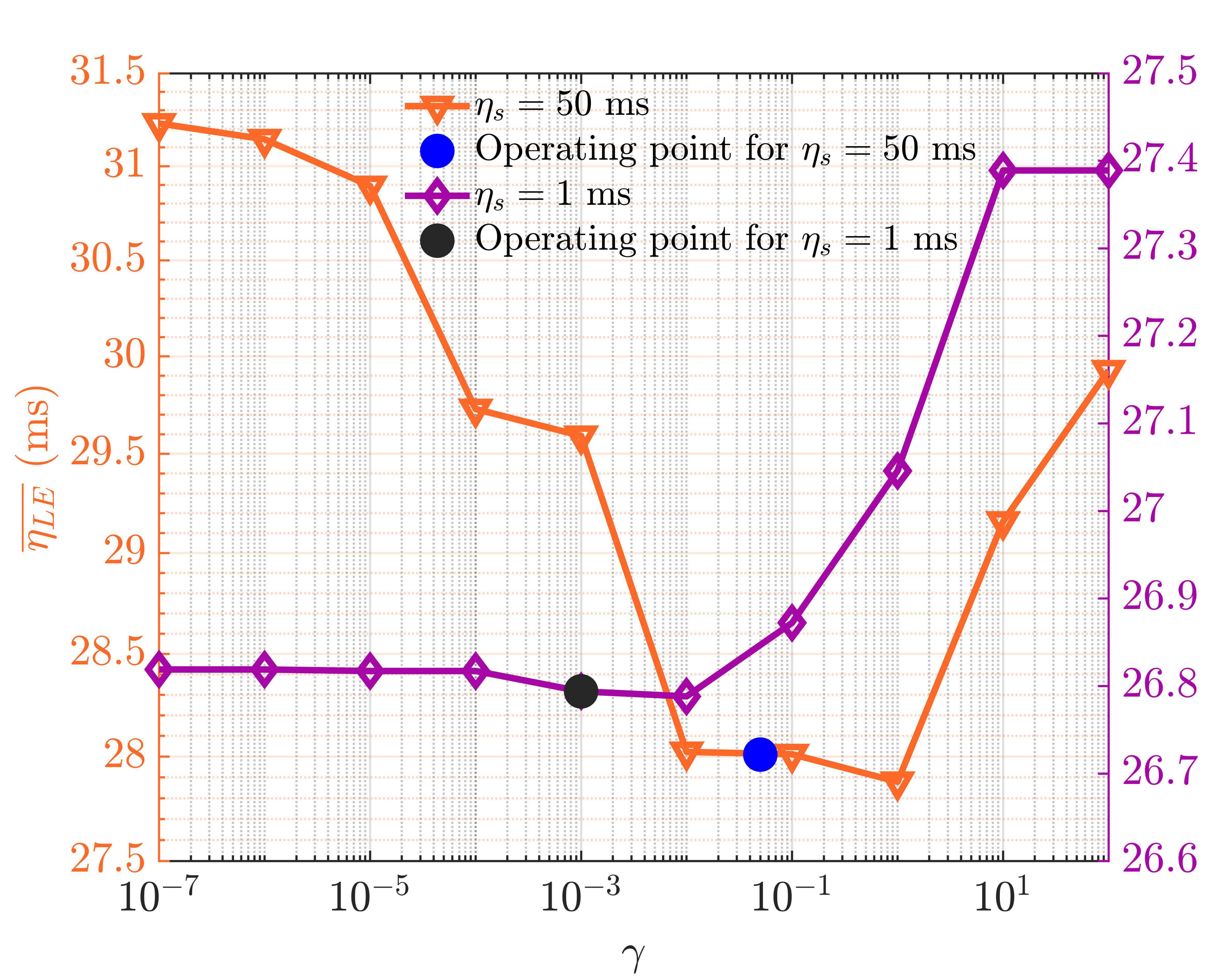}
    % \caption{Average delay component}
    % \label{gammavariation}
    % \vskip\baselineskip
    \caption{Plot of $\overline{\eta_{LE}}$ vs $\gamma$ in ISASR for two different LISL setup delay ($\eta_s$) values along with the operating points for respective $\eta_s$. The orange curve with $\eta_s=50$ ms uses the left y-scale and the purple curve with $\eta_s=1$ ms uses the right y-scale. In both curves, $\overline{\eta_{LE}}$ shows a convex nature with respect to $\eta_s.$}\label{gammavariation}
\end{figure}

In Fig. \ref{gammavariation}, we show how $\overline{\eta_{LE}}$ varies with $\gamma$ in ISASR for New York-London inter-continental connection for a specific $\eta_s$ value. Note that for two different $\eta_s$ values, we use two different y-axis scales for ease of visualization. The purpose is not to compare these two curves with each other, but rather to show that beyond the specific value of $\eta_s$, the $\overline{\eta_{LE}}$ obtained has a convex nature that allows for optimization by fine-tuning the $\gamma$ parameter. When $\gamma$ is smaller than the optimum, from (\ref{modifycost}), we can say that less priority is given to cost$_{st}$ and cost$_{act}$, and more priority to the old cost, i.e., the delay component that leads to unnecessary route change affecting $\overline{\eta_{LE}}$. Similarly, for a higher $\gamma$, more priority is given to cost$_{st}$ and cost$_{act}$ along with less priority to delay component, leading to an increase in the delay component in $\overline{\eta_{LE}}$ and eventually $\overline{\eta_{LE}}$. We also note that the optimal value of $\gamma$ increases with $\eta_s$ which motivates us to select $\gamma$ proportionally to $\eta_s$. This is simply because, with the increase in $\eta_s$, more priority is required to cost$_{st}$ and cost$_{act}$ which also means more priority to the penalty component. We show how close our operating points are to the optimal region. Finding and using the optimal $\gamma$ in ISASR is beyond the scope of this study, and one can easily build a recursive model such as gradient descent to find the optimal $\gamma$.  

\subsection{Time Complexity Analysis}
\begin{figure}[t]
    \centering
    \includegraphics[width=0.48\textwidth]{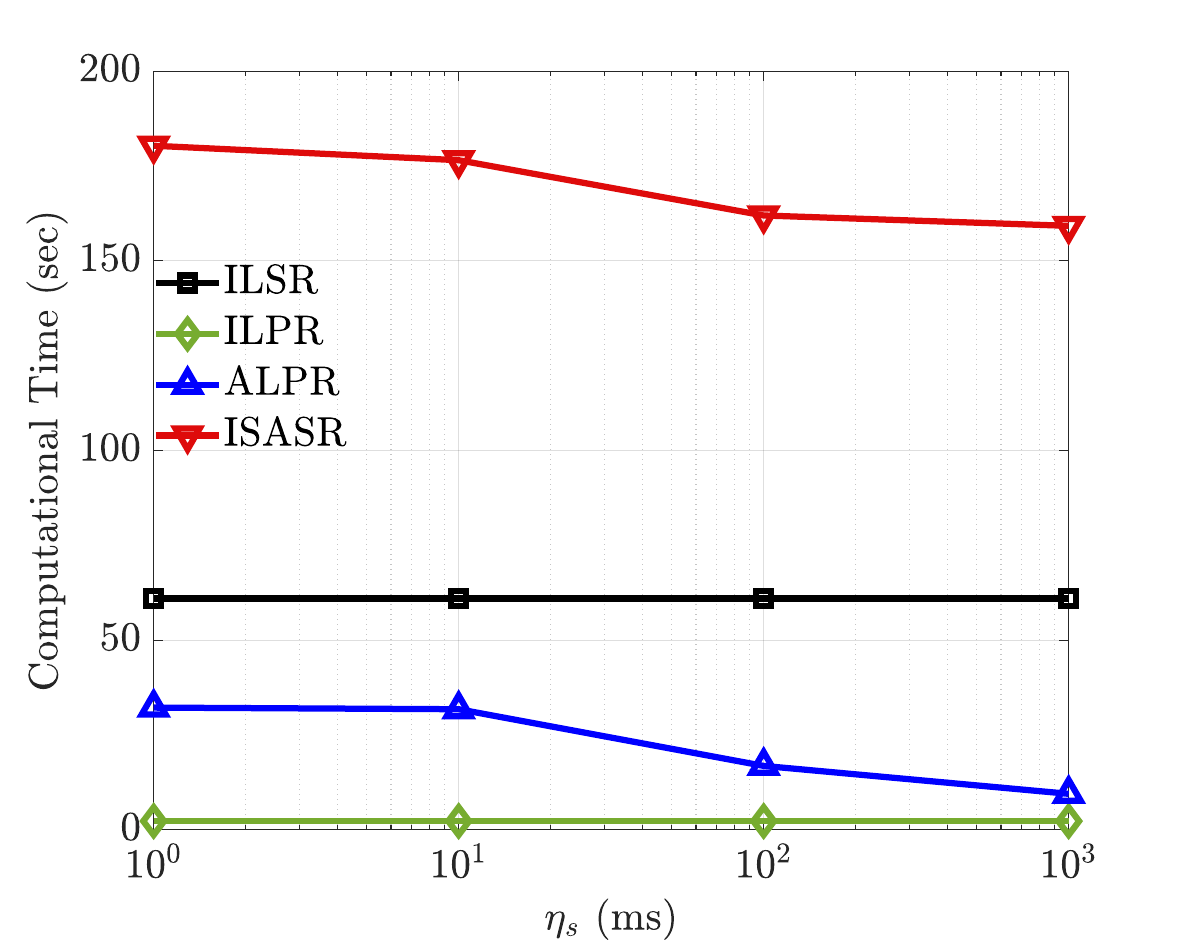}
    \caption{Plot of computational time vs LISL setup delay ($\eta_s$) for New York-London inter-continental connection. Differently than ALPR (shown by solid blue line) and ISASR (shown by solid red line), ILSR (shown by solid black line) and ILPR (shown by solid green line) are not adaptable with $\eta_s$, thus having a constant computational time with respect to $\eta_s$. On the other hand, for ALPR and ISASR, higher $\eta_s$ leads to a lower computational time.}\label{timecomplexity}
\end{figure}

In this discussion, we present a comparative analysis of the computational time for end-to-end route calculations for $600$ time slots. We record the computational time for $100$ iterations and show how the average computational time varies with the LISL setup delay in Fig. \ref{timecomplexity}. In this analysis, computations were performed on a computer equipped with a 2.3 GHz CPU-Intel(R) Core(TM) i5 processor and 20 GB RAM. Clearly, ILPR has the least complexity because of its simplicity. As ILPR only applies DSR when the previous shortest route expires and holds on to the selected route as long as possible, DSR is applied only on the first time slot and whenever the route changes. This makes ILPR the least computationally complex among all three proposed algorithms. In addition, it can be observed that among the three algorithms, the higher the complexity, the better is the performance. ILPR has the least computational time and worst performance compared to ALPR and ISASR. Similarly, ISASR has the highest computational time and best performance. This shows the trade-off between complexity and performance. The rationale behind the constancy of ILSR and ILPR curves with respect to $\eta_s$ is evident. As discussed earlier, for a higher $\eta_s$, ALPR tends to select a route with higher stability, i.e., a route that exists for the longest time. Consequently, if ALPR selects a route with the longest duration of existence, the number of times the algorithm runs to find an end-to-end route for $N$ time slots is less. This leads to a lesser computation time. Conversely, as $\eta_s$ decreases, stability becomes less pivotal, allowing greater emphasis on the delay component. Thus, the selection of the route gradually moves from the highest stable route to the shortest route. So, as $\eta_s$ decreases, the chosen routes exhibit reduced existence times. As a consequence, the number of times the algorithm runs increases and this leads to increase in the computational time. Next, to analyze ISASR, we split Algorithm \ref{eCAR} into three segments that vary with $\eta_s$ as follows: lines 7 to 9 entail the elimination of edges from the search space (computational time$=\tau_1$); line 10 modifies edge costs (computational time$=\tau_2$); and line 12 is to find the shortest route by applying DSR on the network with modified edge costs (computational time$=\tau_3$). It is worth noting that certain lines are not included in $\tau_c,\;c=1,2,3$ because the computational time of those excluded lines does not vary with $\eta_s$. In this comparison study, for four $\eta_s$ values, we keep cost$_{thrsh}$ fixed at $1$ for ISASR. With cost$_{st}$ being directly proportional to $\eta_s$, cost$_{st}$ increases with $\eta_s$, necessitating the elimination of more edges. This contributes to an increase in $\tau_1$. Paradoxically, this has a converse impact on $\tau_2$. As $\eta_s$ increases, more edges are removed from the search space, results in fewer edges necessitating cost modifications, thereby reducing $\tau_2$. In addition, fewer edges exist in the search space, leading to DSR being applied on the truncated network, and consequently reducing $\tau_3$. Given that $\tau_2+\tau_3$ surpasses $\tau_1$, the collective outcome of these factors results in a decline in computational time as $\eta_s$ is increased in ISASR.

\begin{figure*}[]
    % \captionstyle{centerlast}
    \centering
    \begin{minipage}[l]{0.4\textwidth}
        \centering
        \includegraphics[width=\textwidth]{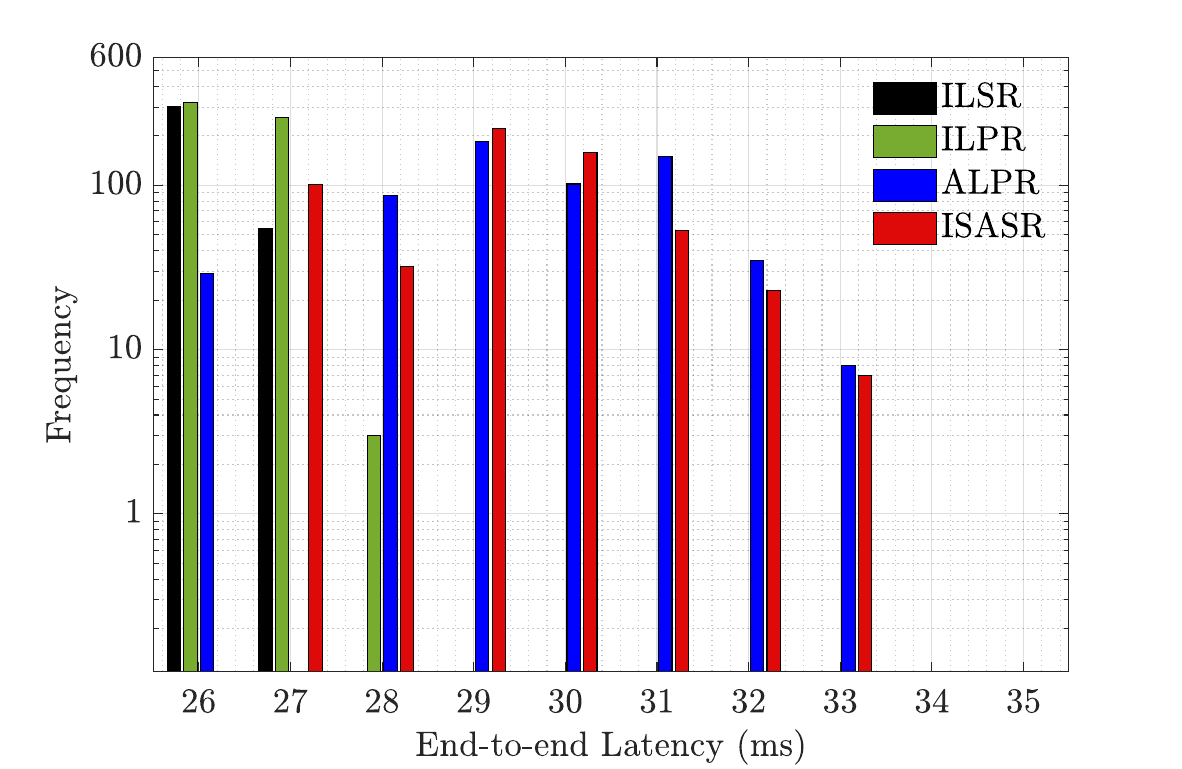}
        % \caption{AAA}\label{fig:AAA}
    \end{minipage}
    \hspace{0.05\textwidth}
    \begin{minipage}[r]{0.4\textwidth}
        \centering
        \includegraphics[width=\textwidth]{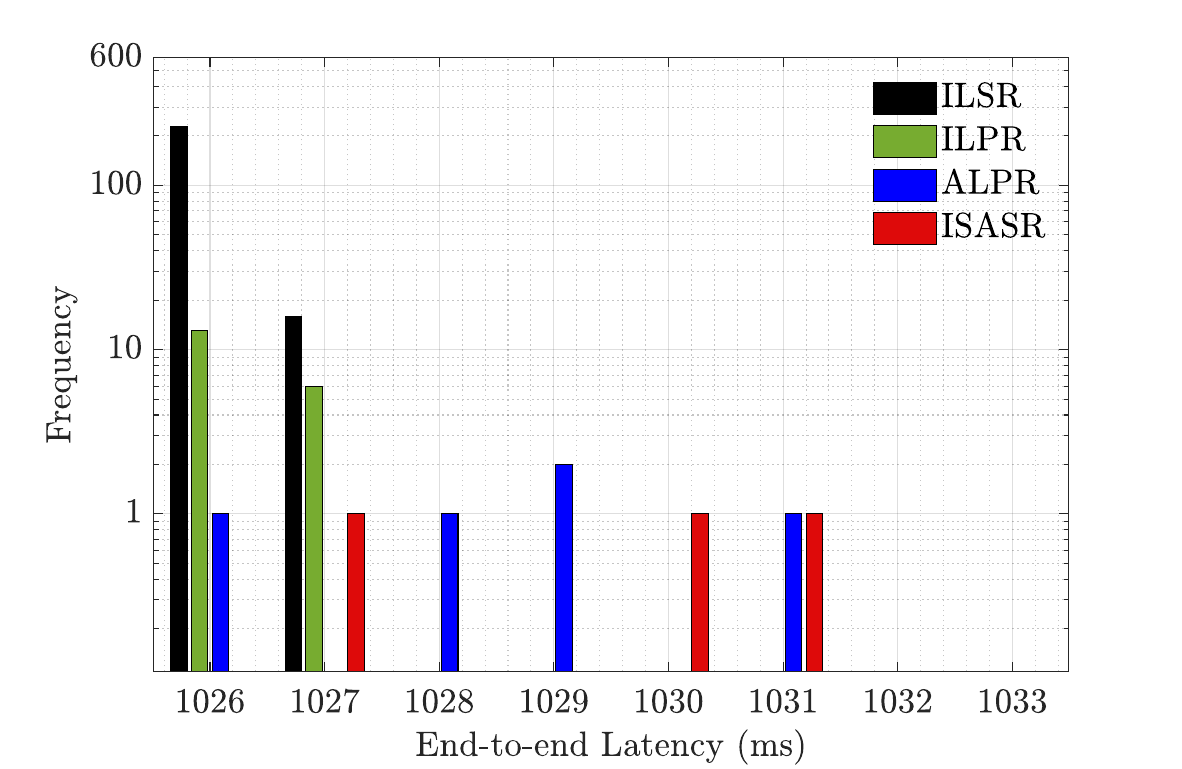}
        % \caption{BBB}\label{fig:BBB}
    \end{minipage}
    \caption{Histogram of instantaneous end-to-end latency with $1000$ ms LISL setup delay.}\label{histogram_1000ms}
\end{figure*}

\begin{figure*}[]
    % \captionstyle{centerlast}
    \centering
    \begin{minipage}[l]{0.4\textwidth}
        \centering
        \includegraphics[width=\textwidth]{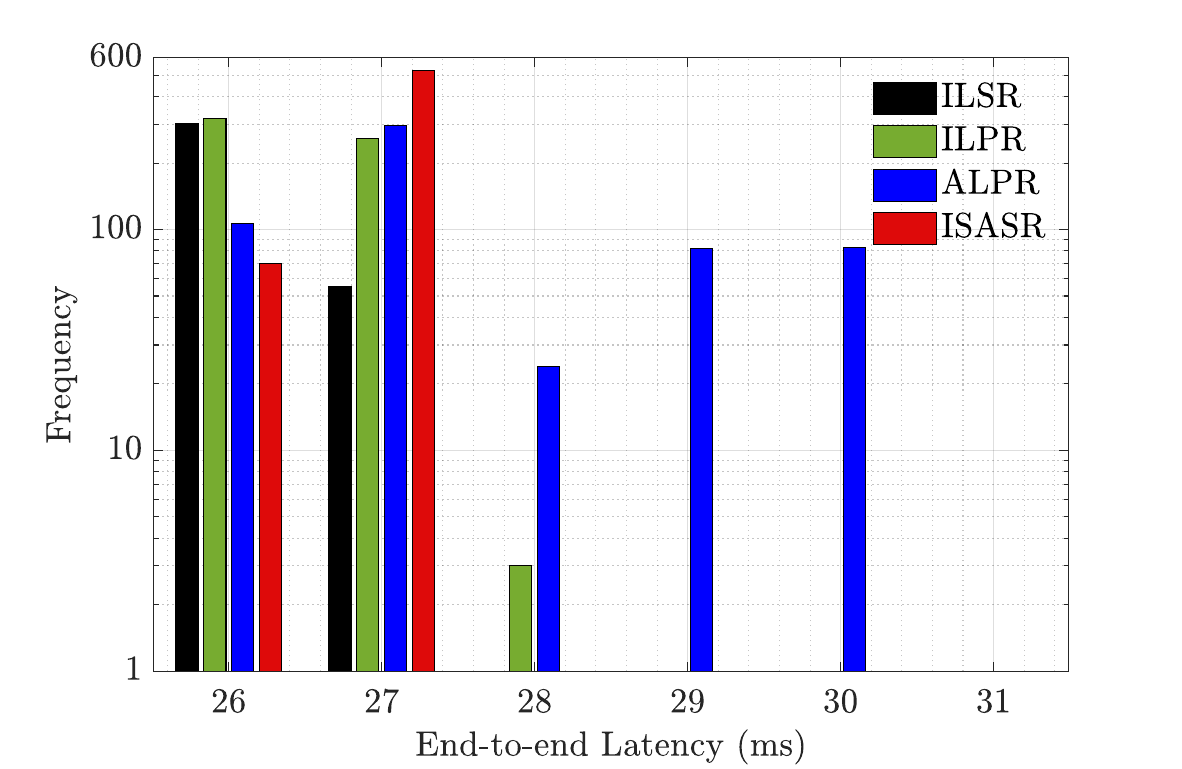}
        % \caption{AAA}\label{fig:AAA}
    \end{minipage}
    \hspace{0.05\textwidth}
    \begin{minipage}[r]{0.4\textwidth}
        \centering
        \includegraphics[width=\textwidth]{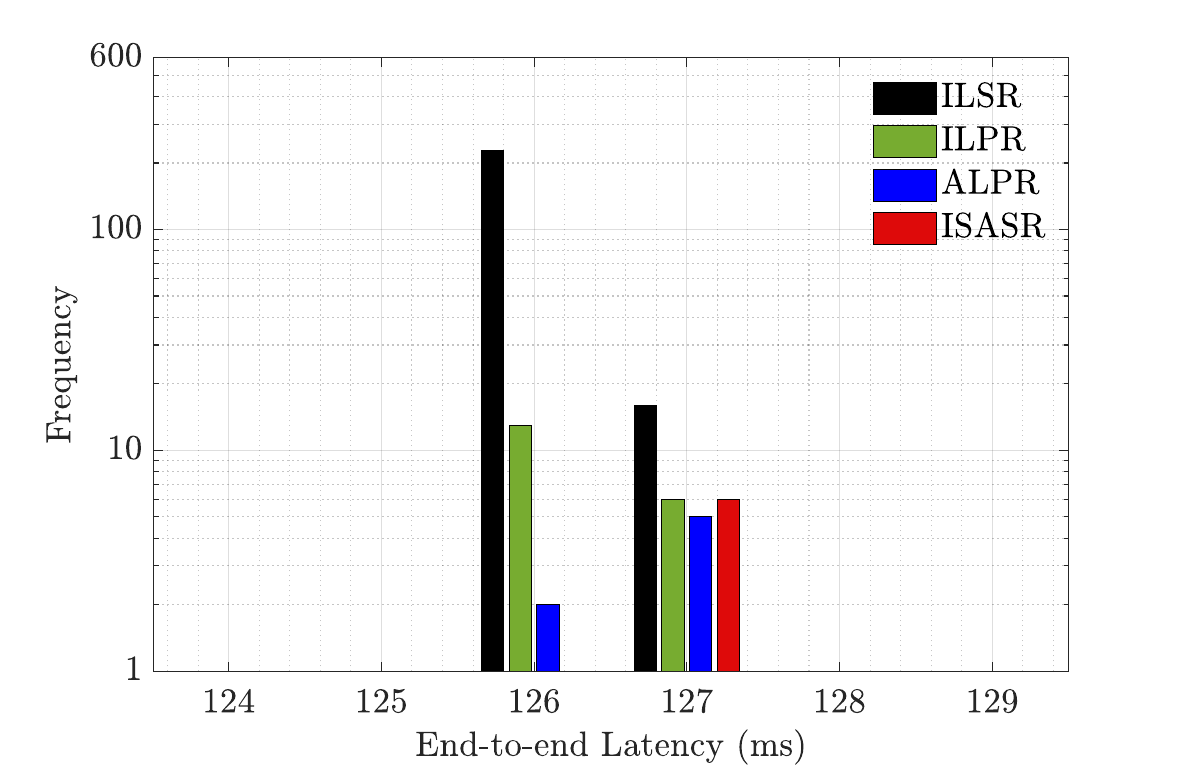}
        % \caption{BBB}\label{fig:BBB}
    \end{minipage}
    \caption{Histogram of instantaneous end-to-end latency with $100$ ms LISL setup delay.}\label{histogram_100ms}
\end{figure*}

\begin{figure*}[]
    % \captionstyle{centerlast}
    \centering
    \begin{minipage}[l]{0.4\textwidth}
        \centering
        \includegraphics[width=\textwidth]{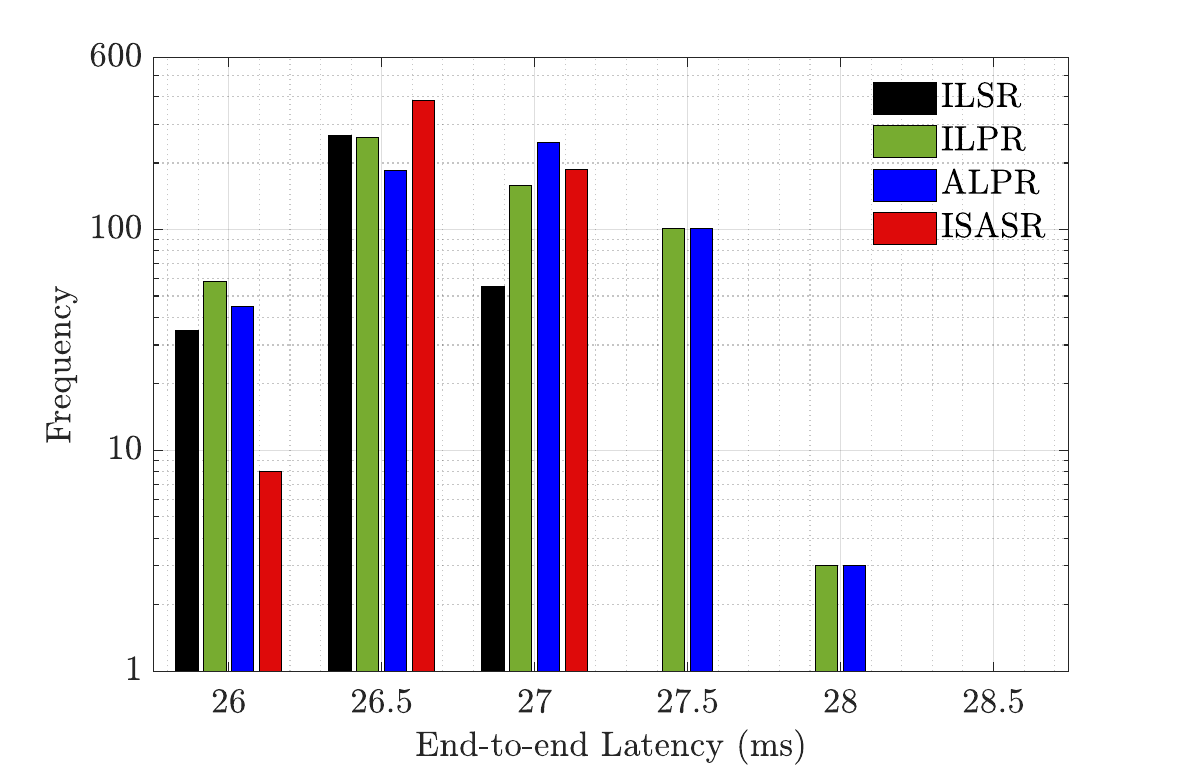}
        % \caption{AAA}\label{fig:AAA}
    \end{minipage}
    \hspace{0.05\textwidth}
    \begin{minipage}[r]{0.4\textwidth}
        \centering
        \includegraphics[width=\textwidth]{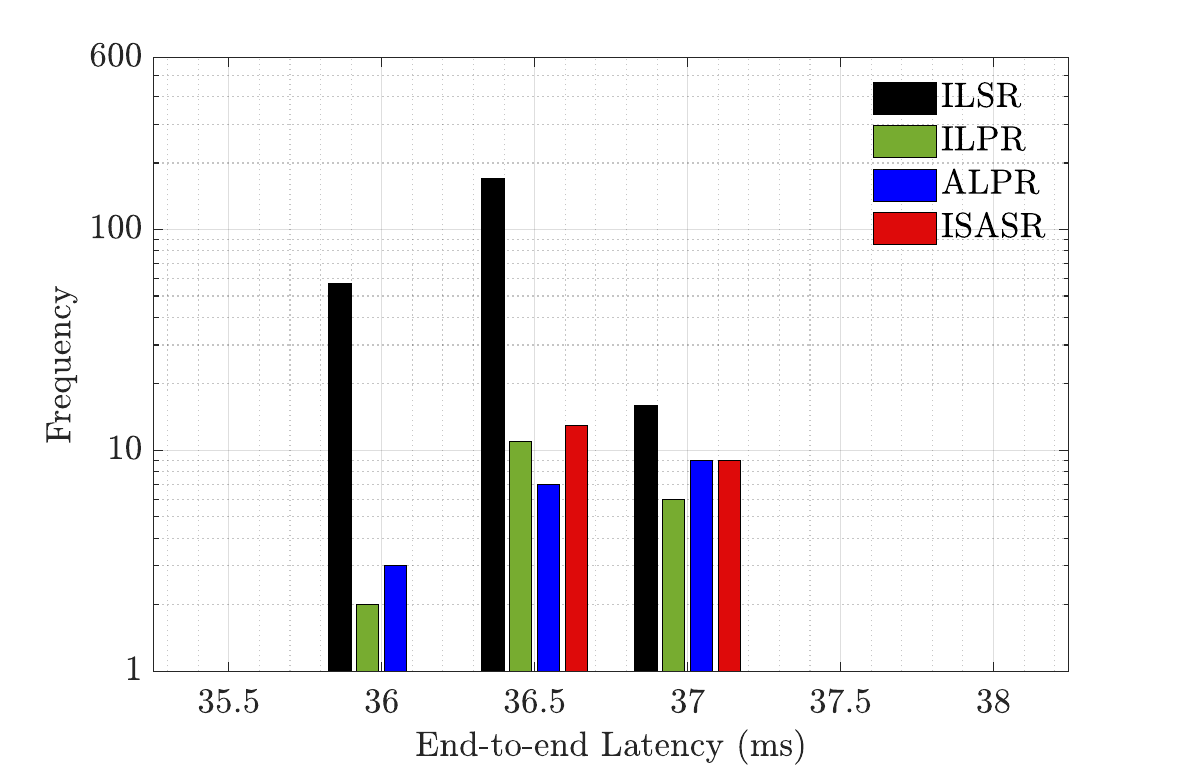}
        % \caption{BBB}\label{fig:BBB}
    \end{minipage}
    \caption{Histogram of instantaneous end-to-end latency with $10$ ms LISL setup delay.}\label{histogram_10ms}
\end{figure*}

\begin{figure*}[]
    \centering
    \begin{minipage}[l]{0.4\textwidth}
        \centering
        \includegraphics[width=\textwidth]{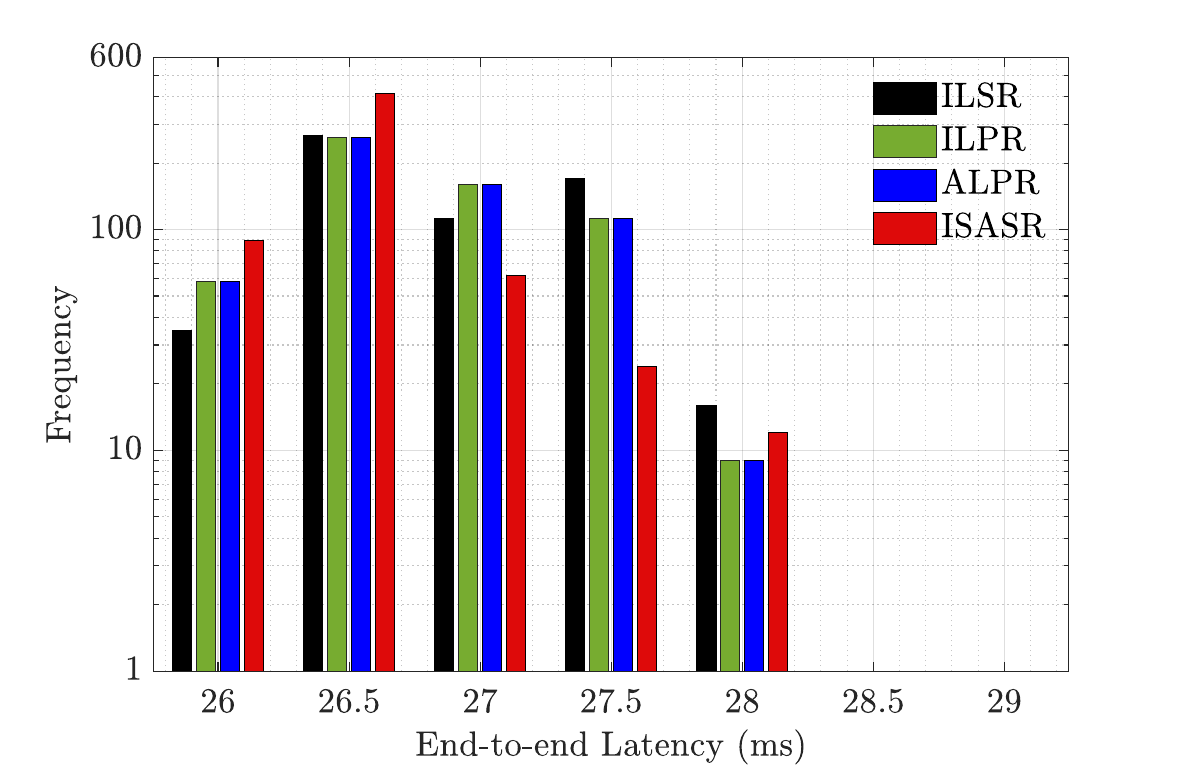}
        \caption{Histogram of instantaneous end-to-end latency with $1$ ms LISL setup delay.}\label{histogram_1ms}
    \end{minipage}
    \hspace{0.05\textwidth}
    \begin{minipage}[r]{0.4\textwidth}
        \centering
        \includegraphics[width=\textwidth]{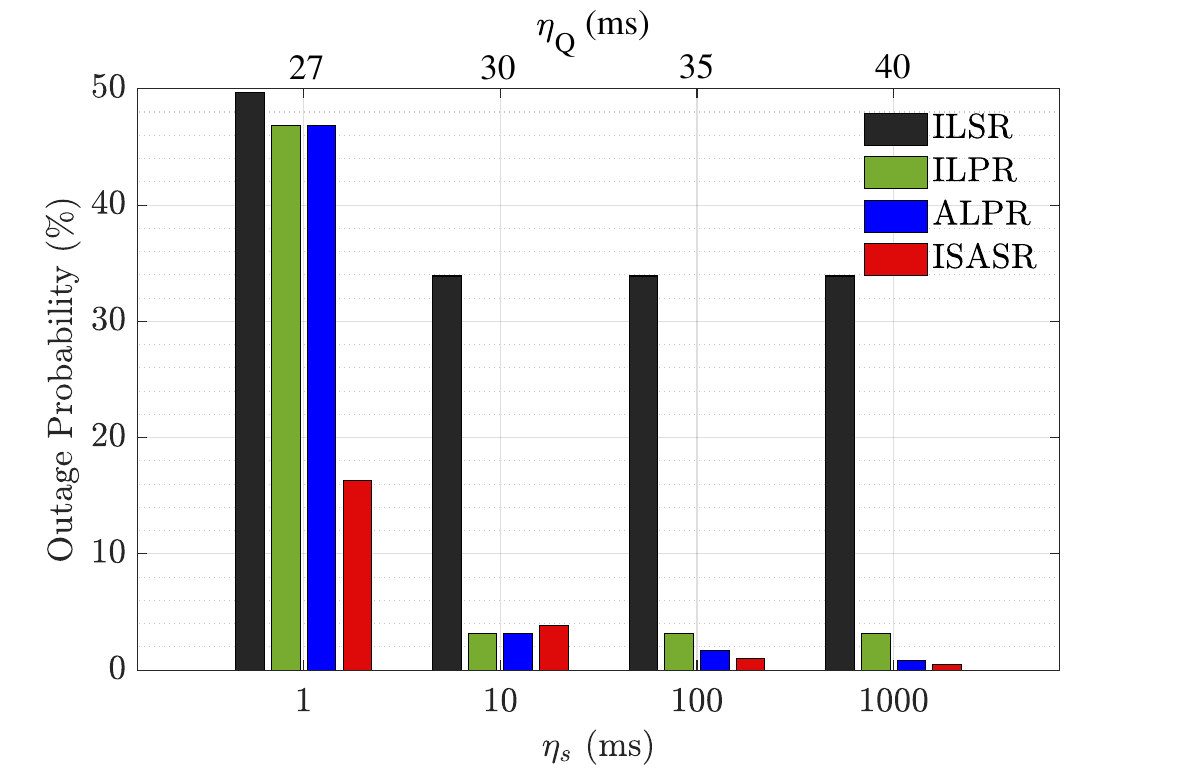}
        \caption{Outage probability vs LISL setup delay.}\label{outage}
    \end{minipage}
\end{figure*}

% \twocolumn
\subsection{Outage Probability}

Until now, we have discussed the performance of the routing algorithms from the perspective of average end-to-end latency. However, from the user experience perspective, along with the average latency performance, worst-case delay is also relevant for acceptable QoS. Although the proposed algorithms are not designed to handle worst-case delays, we compare these algorithms from the outage probability perspective, where an outage event is defined as the instantaneous end-to-end latency being greater than the QoS constraint, i.e., an end-to-end latency threshold. In this context, we first discuss the histogram plots of the instantaneous end-to-end latencies for four $\eta_s$ values in Figs. \ref{histogram_1000ms}, \ref{histogram_100ms}, \ref{histogram_10ms}, and \ref{histogram_1ms}. It is worth highlighting that each of the histograms presented in Figs. \ref{histogram_1000ms}, \ref{histogram_100ms}, and \ref{histogram_10ms} has been separated into two distinct figures. This division is due to a notable gap that exists between two sections within the complete histogram. The portion with lower end-to-end latencies are the occurrences where there is no route change, and higher end-to-end latencies are those with the route change events. In addition, the latency gap between these two portions is in the order of $\eta_s$ value, as expected. In Figs. \ref{histogram_1000ms}, \ref{histogram_100ms}, and \ref{histogram_10ms}, we can observe that there are fewer occurrences of ILPR in the higher latency portions compared to ILSR and more in the lower latency side. This is simply because ILPR has a lower route change rate than ILSR.
As expected, the histograms do not change for ILSR and ILPR with $\eta_s$. On the contrary, in ALPR and ISASR, as shown in Fig. \ref{histogram_1000ms} that most of the occurrences are in the higher side of the low latency portion of the histogram. This is due to the fact that in order to avoid the inclusion of a high penalty, ALPR and ISASR must go through longer and more stable routes in the network. As $\eta_s$ is reduced (Figs. \ref{histogram_100ms}-\ref{histogram_1ms}), this dense occurrence part of ALPR and ISASR in the low latency portion of the histogram gradually shifts left which signifies ALPR and ISASR gradually move on from longer stabler routes to shorter less stable routes, as the inclusion of penalty for establishing a route reduces.

In Fig. \ref{outage}, we show a comparative analysis of outage probability for the four routing algorithms, and for four different $\eta_s$ values. As discussed before, $\eta_s$ may be reduced from tens of seconds to milliseconds, and the QoS constraint will be more stringent in the future. Therefore, we consider a decreasing QoS constraint, $\eta_Q$ as $40, 35, 30,$ and $27$ ms for $\eta_s$ as $1000, 100, 10,$ and $1$ ms, respectively\footnote{QoS constraint for $\eta_s=1$ ms is considered as 27 rather than 25 as the minimum end-to-end delay for New York-London with $1500$ km LISL range is more than 26 ms}. In Fig. \ref{outage}, for $\eta_s=10, 100$, and 1000 ms, outage probabilities are basically represented by respective route change rates. This is because the respective QoS constraints are in between the two portions of the histogram, and the portions with higher end-to-end latencies are due to the route change events. Thus, ILSR and ILPR outage probabilities are constant, and ALPR and ISASR curves are decreasing for those three $\eta_s$ values. This concept is not applicable for $\eta_s=1$ ms scenario as the lower and higher histogram portions overlap with each other in this case.
% Fig. \ref{outage} also signifies the importance of new routing algorithms adaptive with QoS constraints as future research directions.

\subsection{Average Jitter Performance}

\begin{figure}[t]
    \centering
        \includegraphics[width=0.48\textwidth]{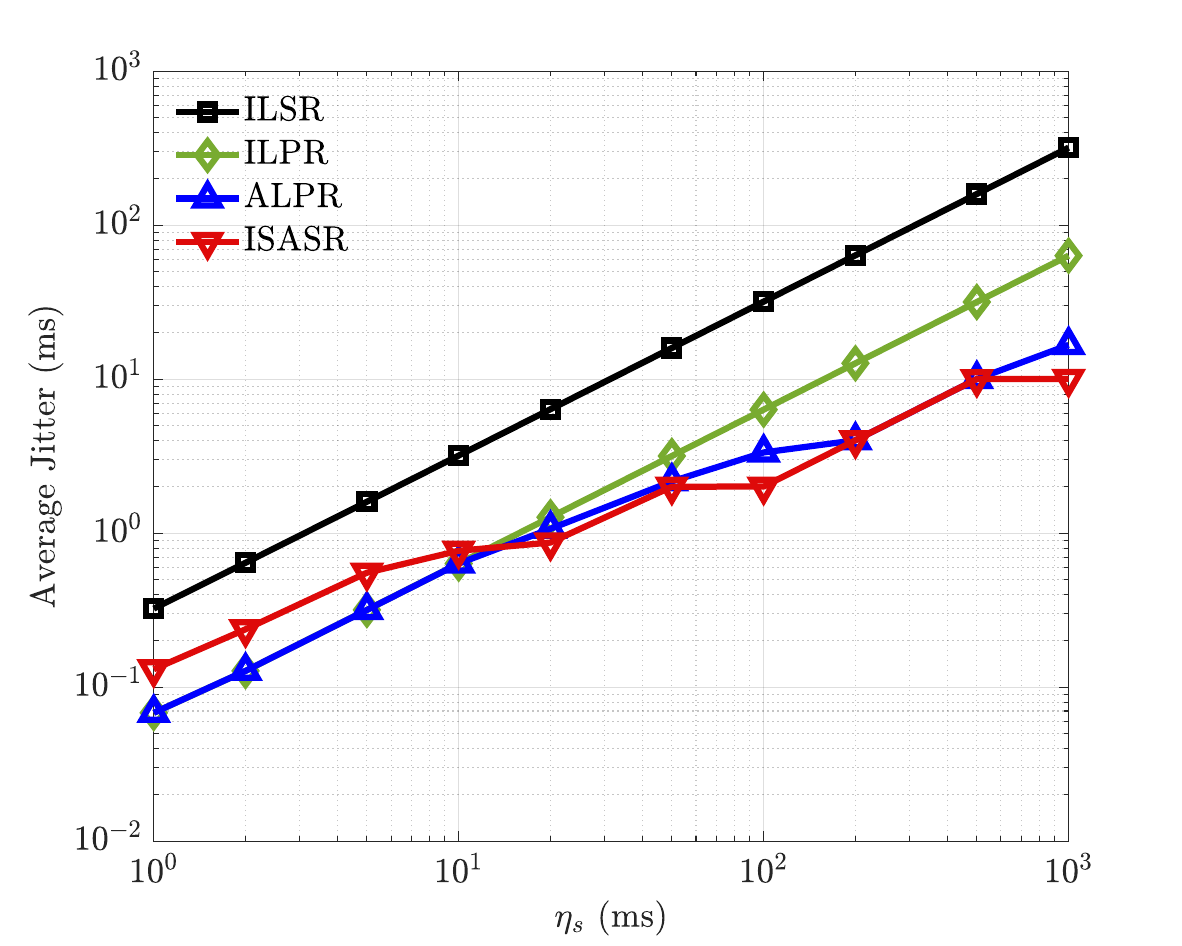}
        % \caption{Average delay component}
        % \label{gammavariation}
       % \vskip\baselineskip
    \caption{Plot of average jitter vs LISL setup delay ($\eta_s$) for New York-London inter-continental connection. The general trend shows that higher $\eta_s$ leads to a higher average jitter for all four algorithms (ILSR, ILPR, ALPR, and ISASR). The average jitter is maximum for ILSR (depicted by the solid black line) for any $\eta_s$. On the other hand, for medium to high values of $\eta_s$, the average jitter is minimum for ISASR (represented by the solid red line), whereas, for low values of $\eta_s$, ALPR (represented by the solid blue line) and ILPR (represented by the solid green line) show the minimum average jitter.}\label{jitterplot}
\end{figure}

In Fig. \ref{jitterplot}, we present the comparative study of average jitter of four routing algorithm varying with $\eta_s$. The average jitter is measured as the average of the differences in the end-to-end latencies of two consecutive time slots, as shown below:
\begin{equation}
    \text{Average Jitter}=\frac{1}{N-1}\sum_{i=1}^{N-1} \big|\delta_r^{[i]}-\delta_r^{[i+1]}\big|.
\end{equation}
As $\eta_s$ increases, the fluctuation in end-to-end latencies increases as well, which eventually leads to an increase in the average jitter. Interestingly, for ILSR and ILPR, the average jitter increases at the same rate, as they are not adaptive to $\eta_s$. On the contrary, as ALPR and ISASR reduce the route change rate with an increase in $\eta_s$, the rate at which the average jitter increases, reduces for them with the increase in $\eta_s$. For the lower range of $\eta_s$, as ILPR and ALPR essentially become the same, their average jitter values also become the same. In the medium to high range of $\eta_s$, ISASR performs the best from the average jitter performance perspective owing to its low path change rate compared to the other three algorithms. On the other hand, as $\eta_s$ is reduced, in ISASR, the end-to-end route changes more frequently than that in ILPR and ALPR. This degrades ISASR performance compared to ILPR and ALPR. 

\section{CONCLUSION AND FUTURE WORK}
\label{sec:conclusion}
%The application of dynamic LISLs for energy-efficient satellite mega-constellation operation in NG and NNG FSOSNs includes the establishment of laser links reactive to traffic demand. This on-demand establishment of LISLs affects end-to-end latencies because of the inclusion of LISL setup delay. In this context, we formulated a mathematical optimization framework to determine the optimum route that minimizes the total end-to-end latencies for one source-destination ground station pair for any given LISL setup delay. Being an ILP problem, we designed three heuristic algorithms, from simple to complex, to have end-to-end routes and compared them with a benchmark algorithm in terms of average end-to-end latency for New York-London and New York-Hanoi inter-continental connection. We also presented a comparative study of the computational time of the proposed and benchmark algorithms and highlighted the trade-off between performance and complexity. Finally, we studied the distributions of instantaneous end-to-end latencies along with outage and average jitter performance, which highlighted the importance of worst-case delay based routing. 

%In this paper, our objective was to find routes based on dynamic topology, i.e., without considering data packets considering only one source destination ground station pair. This framework can be extended to more complex and realistic scenarios, as follows:

Improving the energy efficiency of NG and NNG FSOSNs will require the calculation of on-demand routes and the consequent establishment of dynamic LISLs. This will involve considering link setup delays as penalties that will affect end-to-end latency. In this work, we formulated a mathematical optimization framework to determine the routes that minimize average end-to-end latency for a source-destination ground station pair, and for any LISL setup delay. 

To solve the problem with tractable complexity, we designed three heuristic algorithms with a varying degree of complexity. These algorithms were compared with a benchmark algorithm in terms of average latency in two inter-continental scenarios: New York-London and New York-Hanoi.
 The results show that ILPR, ALPR, and ISASR schemes are able to reduce the average end-to-end latency in exchange for an increase in execution time. In general, this is achieved by selecting routes that may not have the lowest instantaneous latency, but are more stable over time, and incur a lower route change rate and setup delay penalties. In addition to this, the ALPR and ISASR schemes are adaptive to the setup delay value, and can trade off between minimizing instantaneous latency, and incurring a route change, as the setup delay penalty is reduced.
 
Furthermore, we presented a comparative study of the computational time of the proposed and benchmark algorithms, and highlighted the trade-off between performance and complexity. Finally, we studied the distributions of instantaneous end-to-end latencies along with outage and average jitter performance, which emphasized the importance of worst-case delay based routing. 

As future work, we envision the following extensions as compelling lines of research:

\begin{itemize}
    \item Queue status of different satellites associated with respective active LISLs plays a crucial role in selection of routes and eventually the set of active links at a particular time. In addition to propagation, node, and link setup delays, traffic flow can lead to congestion in different parts of the network causing additional delays. Considering queuing delays in satellites along with link setup delay in the route selection may provide additional advantages to those studied in this paper. In particular, we envision the use of machine learning models to predict congestion situations even before they occur, and use these predictions as input to the route selection and link establishment process. In this regard, usage of additional laser terminals can mitigate the affect of LISL setup delay at a cost of size, weight, and power (SWaP) of the payload and communication overhead. This also necessitates the evaluation of payload power consumption. On the other hand, if the link establishment process is done too conservatively, this may lead to decreased energy efficiency. This motivates the study of the energy efficiency-latency trade-off in next-generation satellite mega-constellations.
    %\item The inclusion of data packets in dynamic satellite networks will introduce congestion scenarios at certain satellite nodes. By taking advantage of the predictable satellite network topology and rerouting traffic around the congested satellite nodes, we can carefully select and establish particular links even before they are actually required. This can entirely eliminate the impact of the LISL setup delay regardless of how high the value is. This can be achieved by proactive algorithms (keeping track of the congestion states of satellites), reactive machine learning algorithms (predicting congestion), or hybrid approaches. On the contrary, as extra links have to be established prior to when they are actually needed, it will increase the overall network energy consumption. This motivates us to study  the power-delay trade-off characteristics of next-generation satellite mega-constellations.
    \item In the next 10-20 years, there will be satellite constellations of different generations, companies, and agencies, with different PAT technology. This implies that LISL configuration delays may be highly heterogeneous. Furthermore, depending on the relative velocities of two satellites, which may be in different orbits, the LISL configuration delay will vary. This indicates the importance of having efficient routing algorithms that can handle heterogeneous LISL setup delays across the satellite network.
    %\item In the next 10-20 years, there will be satellites in the sky from different generations with different levels of PAT technology. This implies that the LISL setup delays will be different for different satellites. In addition, depending on the relative velocities of two satellites, the LISL setup delay will vary. This indicates the importance of efficient routing algorithms that can handle heterogeneous LISL setup delays across the satellite network.
    \item This work focused on the establishment of a route between a pair of terrestrial endpoints. However, the inclusion of different types of traffic (voice, data, video), link and node capacities, and required reliability, may demand the use of more than one route. In this sense, it is also necessary to consider the cases of one source with multiple destinations (site diversity), and multiple sources with multiple destinations having a different traffic load for each source-destination pair.  
    %\item There could be a need for multiple routes from a source to a destination ground station, either due to higher traffic demand than the link capacity that one route is not sufficient or usage of different routes for different types of traffic (e.g., video, voice, data etc.). In addition, instead of one source-destination pair, there could be one source multiple destinations (to employ site diversity), multiple sources one destination, or multiple sources and destinations. Routing in such heterogeneous scenarios is necessary to explore for future mega-constellations.
    \item While the primary goal of this work was to minimize average latency, our findings indicate that worst-case latency and jitter are equally crucial performance indicators for ensuring a smooth user experience. Designing algorithms that optimize these metrics will be the central focus of our future research.
    %\item Although in this paper, our objective is to minimize the average latency, we have shown that worst-case latency and latency jitter are also key performance indicators for a smooth user experience which will be the focus of the future work.  
    \item \db{Finally, to ensure scalability in implementation along with performance improvement, exploring alternative strategies for identifying a finite set of routes, rather than solely considering disjoint routes, could be advantageous.}
\end{itemize}

\bibliographystyle{IEEEtran}
\footnotesize
% \vspace{-0.5em}
\bibliography{IEEEabrv}
\end{document}